\documentclass[journal]{IEEEtran}
\usepackage{graphicx}
\usepackage{cite}
\usepackage{amsmath}
\usepackage{url}
\usepackage{flushend}
\usepackage[latin1]{inputenc}
\usepackage{colortbl}
\usepackage{soul}
\usepackage{multirow}
\usepackage{pifont}
\usepackage{color}
\usepackage{epstopdf}
\usepackage{subfigure}

\begin{document}
	\title{	A Soft-switched Fast Cell-to-Cell Voltage Equalizer for Electrochemical Energy Storage}
	\author{Shimul K Dam and Vinod John\\
		Department of Electrical Engineering,\\
		Indian Institute of Science, Bangalore-560012.\\
		(email: shimuld@iisc.ac.in, vjohn@iisc.ac.in)}
	
	%\author{
	%	\vskip 1em
	%	{\color{red}
	%	First A. Author1, \emph{Student Membership},
	%	Second B. Author2, \emph{Membership},
	%	\\ and Third C. Author3, \emph{Membership}
	%	}
	%
	%	\thanks{
	%		
	%		{\color{red}
	%		Manuscript received Month xx, 2xxx; revised Month xx, xxxx; accepted Month x, xxxx.
	%		This work was supported in part by the xxx Department of xxx under Grant  (sponsor and financial support acknowledgment goes here).
	%		
	%		(Authors' names and affiliation) First A. Author1 and Second B. Author2 are with the xxx Department, University of xxx, City, State C.P. Country, on leave from the National Institute for xxx, City, Country (e-mail: author@domain.com). 
	%		
	%		Third C. Author3 is with the National Institute of xxx, City, State C.P. Country (corresponding author to provide phone: xxx-xxx-xxxx; fax: xxx-xxx-xxxx; e-mail: author@ domain.gov).
	%		}
	%	}
	%}
	
	\maketitle
	
	\begin{abstract}
		Batteries are connected in series to meet the voltage requirement in many applications. A voltage equalizer circuit is necessary to ensure that none of the batteries is over-charged or over-discharged.	
		A novel fast soft-switched cell-to-cell voltage equalizer topology is proposed in this work. This topology can transfer charge from multiple over-charged batteries to multiple under-charged batteries simultaneously avoiding any unnecessary charging or discharging of a battery to achieve fast voltage equalization. The proposed circuit topology and modulation method ensure zero voltage switching under all battery conditions. The circuit operation and soft-switching are analyzed and experimentally verified with a four battery voltage equalizer prototype. The prototype is tested with a battery bank and a hybrid ultra-capacitor bank, and a high conversion efficiency is verified. 
	\end{abstract}
	
	
	%\markboth{IEEE TRANSACTIONS ON INDUSTRIAL ELECTRONICS}%
	{}
	
	\definecolor{limegreen}{rgb}{0.2, 0.8, 0.2}
	\definecolor{forestgreen}{rgb}{0.13, 0.55, 0.13}
	\definecolor{greenhtml}{rgb}{0.0, 0.5, 0.0}
	
	\section{Introduction}
	
	\IEEEPARstart{E}{Electro-chemical} energy storage plays an essential role in electric vehicles, power backup systems, renewable energy generation systems, and many other applications. Batteries are widely used as energy storage. Often a number of batteries are connected in series to get a higher terminal voltage which suits the application. All series-connected batteries are charged and discharged together. Although the charging and discharging current are same for all the batteries, their terminal voltages may not be the same. This is a result of unequal charge capacities of the battery due to manufacturing tolerance, unequal aging, and non-uniform temperature distribution. Hence, the batteries with lower charge capacity experience over-charge and over-discharge in each charge-discharge cycle. This over-charge and over-discharge further reduces their charge capacities escalating the problem. If this situation is allowed to continue, some of the batteries will fail much earlier than their expected lifespan. Hence, a voltage equalizer circuit is necessary to ensure equal voltages of all the batteries by redistributing charge among them. This circuit takes charge from any over-charged battery and gives it to any under-charged battery.

	There are many voltage equalizer topologies proposed in the literature. Ideally, a voltage equalizer should take charge only from the over-charged batteries and transfer this charge only to the under-charged batteries without disturbing the rest of the batteries. The cheapest and simplest topology is a dissipative one where an overcharged battery is discharged through a resistor to equalize voltages of the batteries\cite{14}. This approach is not a practical one for a large battery bank since a lot of energy is wasted in the process. Hence, all active voltage equalizer topologies use power converters to transfer charge with high efficiency. Some of these converter based topologies transfer power from an individual battery to the entire battery stack. This task is achieved by different approaches such as multiple transformer based converter\cite{19}, single transformer with multiple secondary windings\cite{30,ein,lim,hua3,hua1,hua2}, single dc-dc converter with selection switches\cite{16,6,8,10,zhang2,hannan}, inverter with voltage multiplier circuit\cite{9,24} etc. These topologies perform charge transfer between individual battery and the battery string. But, this approach leads to unnecessary charging or discharging of some of the batteries in the battery string and causes delay in equalization and extra power loss. The adjacent cell topologies \cite{15,yua,hua4,27,28, 29} equalizes voltage of two adjacent batteries. This adjacent battery voltage equalization is done for each pair of adjacent batteries. This approach is simpler and does not require voltage monitoring, but charge transfer between two non-adjacent batteries has to be through all the batteries between them. Hence, this approach also leads to unnecessary charging or discharging of some of the batteries causing extra delay and power loss. 
	
	The unnecessary charging and discharging can be avoided if the voltage equalizer can directly transfer charge from over-charged batteries to under-charged batteries. Very few works have been reported which can accomplish this. A selection switch network and isolated power converter can be used to select one overcharged battery and one under-charged battery and transfer charge between them\cite{yar,32,xiong, lee}. Although this approach uses less number of high frequency switches and achieves direct cell-to-cell charge transfer, batteries are selected sequentially and voltage equalization happens between a pair of batteries at a time. Hence, this process takes a long time. The fastest way to equalize the voltages of a string of series connected batteries is to transfer charge from all over-charged batteries to all under-charged batteries simultaneously without charging or discharging the rest of the batteries in the string. A multi-winding transformer based approach is proposed in\cite{shang1} to achieve multi-cell to multi-cell charge transfer which suffers from the problem of leakage inductance of the transformer windings and difficulties in implementation of the multi-winding transformer. Multi-cell to multi-cell charge transfer is achieved using switched capacitors in \cite{shang2} and \cite{ye}. A delta-structured switched capacitor network in used in \cite{shang2} where the number of capacitors increases rapidly with the number of batteries in series. The topology proposed in \cite{ye} is simpler and uses only one capacitor per battery. The equalization currents in the batteries in the   topologies in \cite{shang1,shang2,ye}  depend on the voltage differences among the batteries. Hence, the equalization power of the equalizer decreases with the progress of the voltage equalization process leading to longer equalization time.
	
	 A bi-directional Cuk converter based topology is proposed in \cite{33} where each battery is connected to a common energy bus through a bi-directional Cuk converter, leading to complex implementation. It requires close-loop control of the battery currents making this approach costly and more complicated to implement. However, the battery currents do not depend on the voltage differences and the equalizer can work with its rated power all the time to achieve faster voltage equalization compared to \cite{shang1,shang2,ye}.  A simpler topology is proposed in\cite{itec} using one half-bridge leg, one capacitor and one inductor for each battery. The battery currents in this topology do not depend on the voltage differences and there is no need for close-loop current control. Thus, the topology proposed in \cite{itec} achieves faster multi-cell to multi-cell charge transfer compared to \cite{shang1,shang2,ye} with a simpler control and circuit implementation compared to \cite{33}.

			  However, the sin-triangle modulation used in \cite{itec} leads to high switching frequency and hard switching, thus reducing converter efficiency.
	
	This work is an improvement over the approach presented in \cite{itec}.
	The topology presented here has structural similarity to \cite{itec}, but with suitable modification, component selection, and modulation method, leads to lower size and cost, simpler control and higher efficiency due to ZVS of all the switches while achieving all the advantages of \cite{itec}. This voltage equalizer can transfer charge from multiple batteries to multiple batteries simultaneously while avoiding unnecessary charging or discharging of any battery. The equalization current does not depend on the voltage differences of the batteries. All these features leads to fast voltage equalization. An open loop control approach with only battery voltage sensing is sufficient to achieve cell balancing.  The proposed modulation helps to achieve soft-switching of the switching devices under all battery conditions leading to high efficiency.

	\section{Topology}%%%%%%%%%%%%%%%%%%%%%%%%%%%%%%%%%%%%%%%%%%%%%%%%%%%%%%%%%%%%%%%%%%%%%%%%%%%%%%%%%%%%%%%%%%%%%%%%%%%%%%%%%%%%%%%%%%%%%%%%%5
\begin{figure}[h!]
		\centering
		\includegraphics[width=6cm]{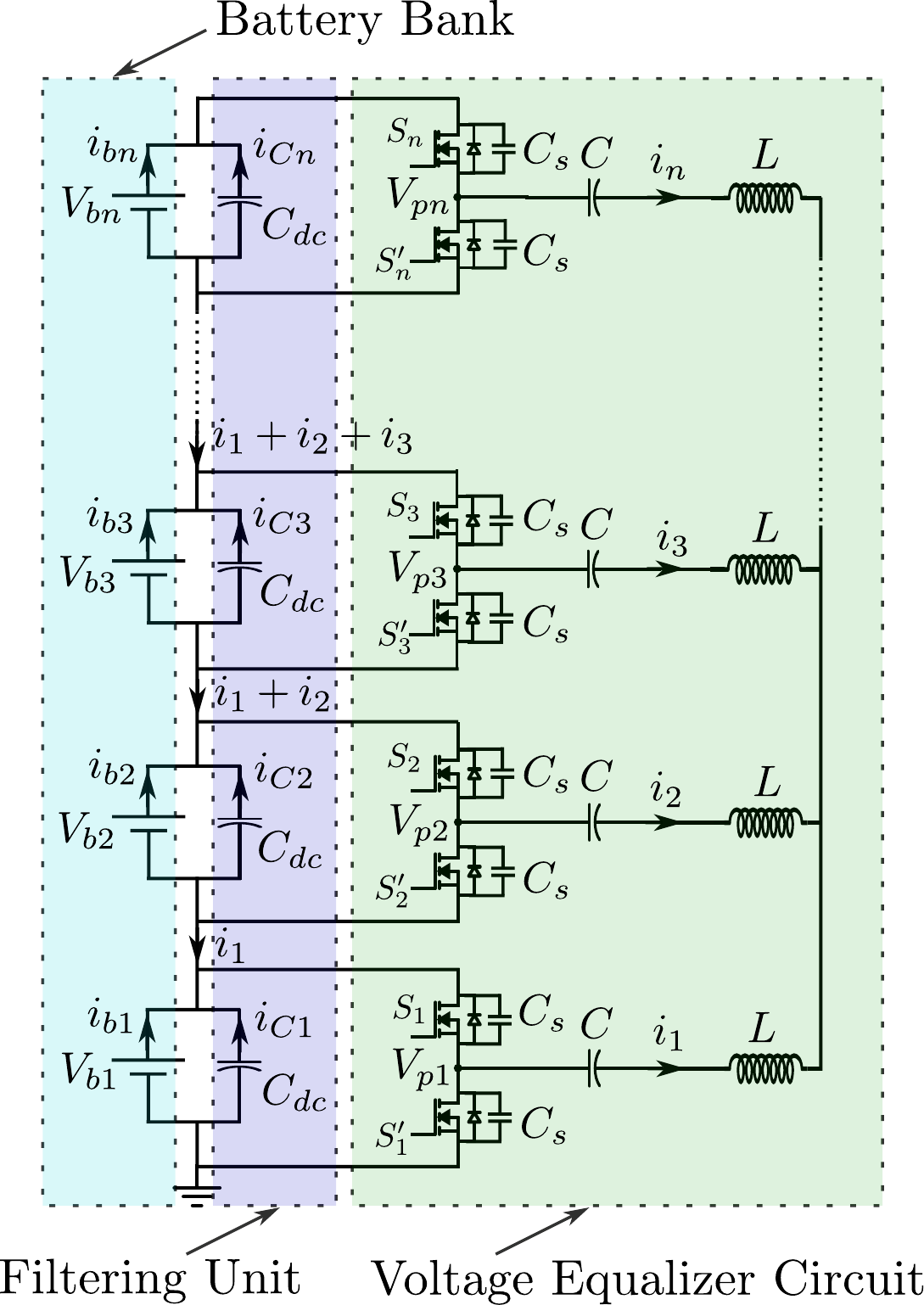}
		\caption{Schematic diagram of the proposed voltage equalizer.}
		\label{schematic}
	\end{figure}

	The circuit topology is adopted from \cite{itec} with an additional small capacitor $C_s$ connected across each switch to achieve ZVS operation as shown in Fig.\,\ref{schematic}. 
 In this topology, each battery has one half-bridge converter leg. The pole of the converter leg is connected to a series combination of a capacitor $C$ and an inductor $L$. One terminal of each inductor is connected at a common node. The capacitors, $C$, connected to the converter poles are used to block dc voltage. The capacitance of $C$ is chosen to be large so that it offers negligible impedance at the switching frequency. The inductor determine the maximum possible current in the converter legs.
 
  The dc bus capacitors, $C_{dc}$ is used as a filtering element to filter out the switching components from the battery current. The capacitor $C_s$ is only used to achieve resonant transition and it has to store a small amount of energy. Hence, a low cost ceramic capacitor with small capacitance value can be used as $C_s$. 
	
	\begin{figure}[h!]
		\centering
		\includegraphics[width=8cm]{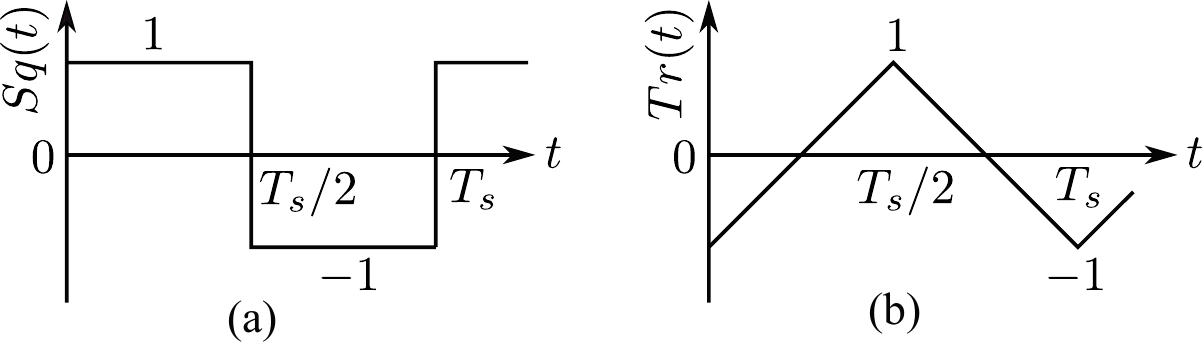}
		\caption{(a) Square wave function, $Sq(t)$, (b) Triangular wave function, $Tr(t)$ .}
		\label{func}
	\end{figure}
	
	\subsection{Modulation}\label{modulation}
	The modulation strategy uses phase shifted square-waves instead of sine-triangle PWM used in \cite{itec}. This modulation technique helps to achieve higher efficiency with lower circuit component requirements and lower complexity in implementing the control algorithm in digital controller. The proposed modulation strategy along with the control algorithm is explained below.
	
	The top device of each half bridge converter leg is controlled with a square wave of $0.5$ duty and time period $T_s$. However, the phase of each of these square waves is determined based on the condition of the corresponding battery. For describing these signals and subsequent analysis, two periodic functions $Sq(t)$ and $Tr(t)$ of time period $T_s$ are defined as shown in Fig.\,\ref{func}. An Op-Amp based non-isolated voltage sensor is used to sense all the battery voltages. The sensed voltages are then sent to the digital controller using Analog to Digital Converter (ADC). The digital controller then decides which battery voltages are within acceptable voltage range, which batteries have to be charged and which ones to be discharged. 
	
		\begin{figure}[h!]
			\centering
			\includegraphics[width=8cm]{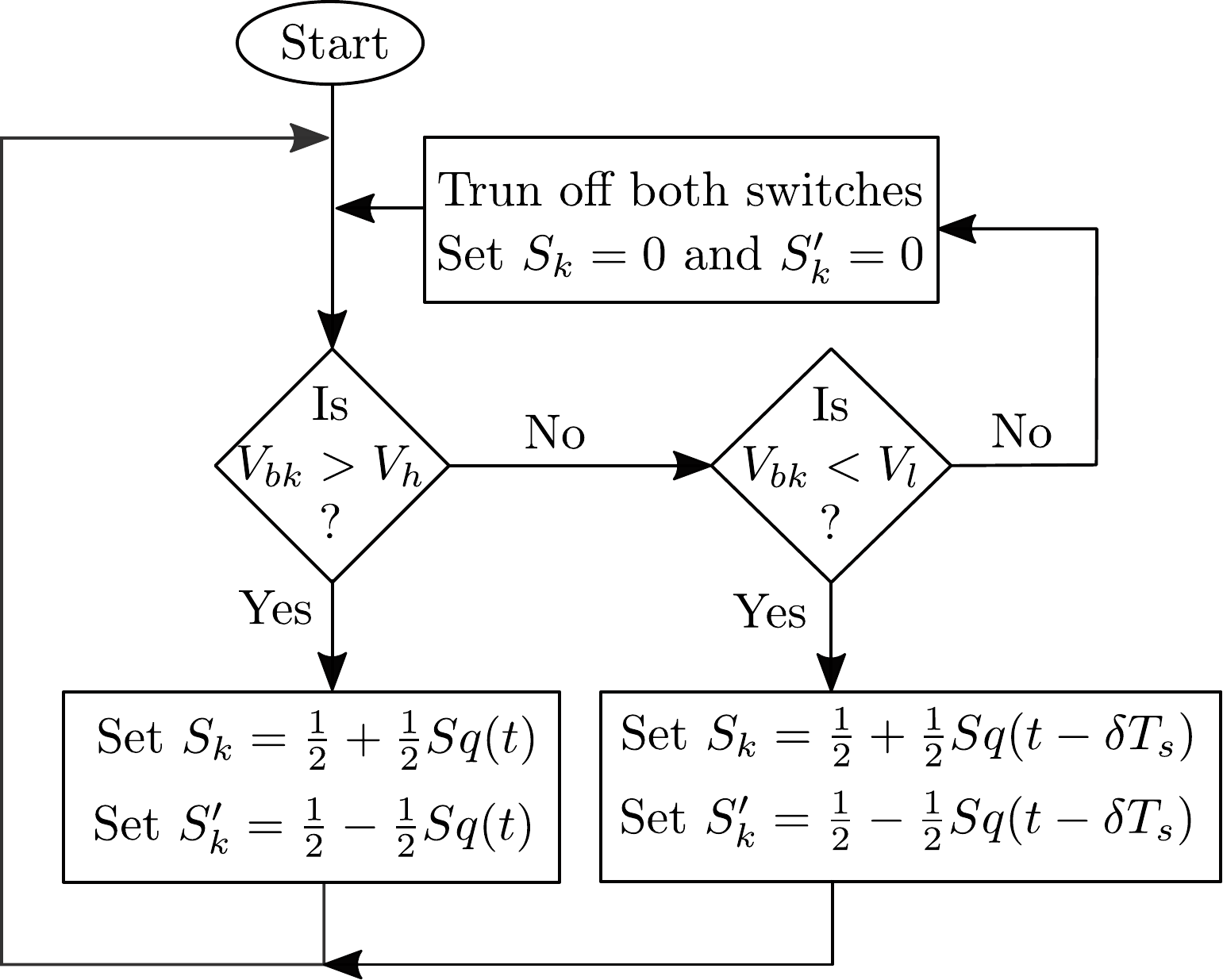}
			\caption{Flowchart of the algorithm for controlling charging and discharging the $k^{th}$ battery.}
			\label{flowchart}
		\end{figure}

	A battery voltage is assumed to be acceptable if it is within a small voltage tolerance $V_{tol}$ around the average voltage of all the batteries. The average voltage of all the batteries for a battery bank consisting of $n$ number of batteries in series is given by,
	\begin{eqnarray}
	V_{avg}=\frac{\sum_{k=1}^{n} V_{bk}}{n}
	\end{eqnarray} 
	Where, $V_{bk}$ is the voltage of the $k^{th}$ battery. The upper limit $V_h$ and the lower limit $V_l$ of the acceptable voltage band around the average voltage are given by,
	\begin{eqnarray}
	V_{h}=V_{avg}+V_{tol}\\
	V_{l}=V_{avg}-V_{tol}
	\end{eqnarray} 
	The decision to charge of discharge a battery is taken based on its voltage based on the algorithm shown in the flowchart in Fig.\,\ref{flowchart} where $S_k$ and $S'_k$ are the gate drive signals respectively of the top and the bottom devices of $k^{th}$ converter leg and $\delta T_s$ is the phase difference among the charging and discharging converters. 
	 
	In general, the gate drive signal for the top device of the $k^{th}$ converter is given by,
	\begin{eqnarray}
		S_{k}=\frac{1}{2}+\frac{1}{2}Sq(t-\delta_k  T_s)
	\end{eqnarray}
	The modulation strategy is as follows,
	\begin{itemize}
		\item If $k^{th}$ battery needs to be discharged then $\delta_k=0$.
		\item If $k^{th}$ battery needs to be charged then $\delta_k=delta$ where $\delta>0$.
		\item For the rest of the batteries, both the devices of the converter leg are turned off.
	\end{itemize}

%	The gate signals for all discharging batteries are the same and the gate signal of the top device is given by, 
%	\begin{eqnarray}
%	\label{d_dis}
%	S_{dis}=\frac{1}{2}+\frac{1}{2}Sq(t)
%	\end{eqnarray}
%	The gate signals for all charging batteries are also same and gate signal for the top device is given by, 
%	\begin{eqnarray}
%	\label{d_ch}
%	S_{ch}=\frac{1}{2}+\frac{1}{2}Sq(t-\delta T_s)
%	\end{eqnarray}
%	For all the batteries which are not to be charged or discharged, both the switches in the converter leg are turned off. Hence, these batteries do not take part in power transfer. The condition for power flow from the discharging batteries to charging batteries is given by,
%	$$\delta>0$$.
	
	 The following subsections discuss how this modulation strategy achieves the goal of voltage equalization among multiple series connected batteries.

	\subsection{Operation}%%%%%%%%%%%%%%%%%%%%%%%%%%%%%%%%%%%%%%%%%%%%%%%%%%%%%%%%%%%%%%%%%%%%%%%%%%%%%%%%%%%%%%%%%%%%%%%%%%%%%%%%%%%%%%%%%%%%%%%%%%%%%%%%%%%%%%%%%%%%%%%%%%%%
	The operation of the equalizer circuit is explained here with equivalent circuit and mode diagrams for a specific case of four batteries where battery 1 and 2 are discharging and battery 3 and 4 are charging. For $k^{th}$ converter in Fig.\,\ref{schematic}, the pole voltage with respect to the negative terminal of $k^{th}$ battery is given by,
	\begin{eqnarray}
		V_{pk}&=&S_k V_{bk}\\
		\label{vpk1}
		&=& \frac{V_{bk}}{2}+\frac{V_{bk}}{2}Sq(t+\delta_k T_s)
	\end{eqnarray}
	
		\begin{figure}[h!]
			\centering
			\begin{subfigure}[]{\includegraphics[width=4.7cm]{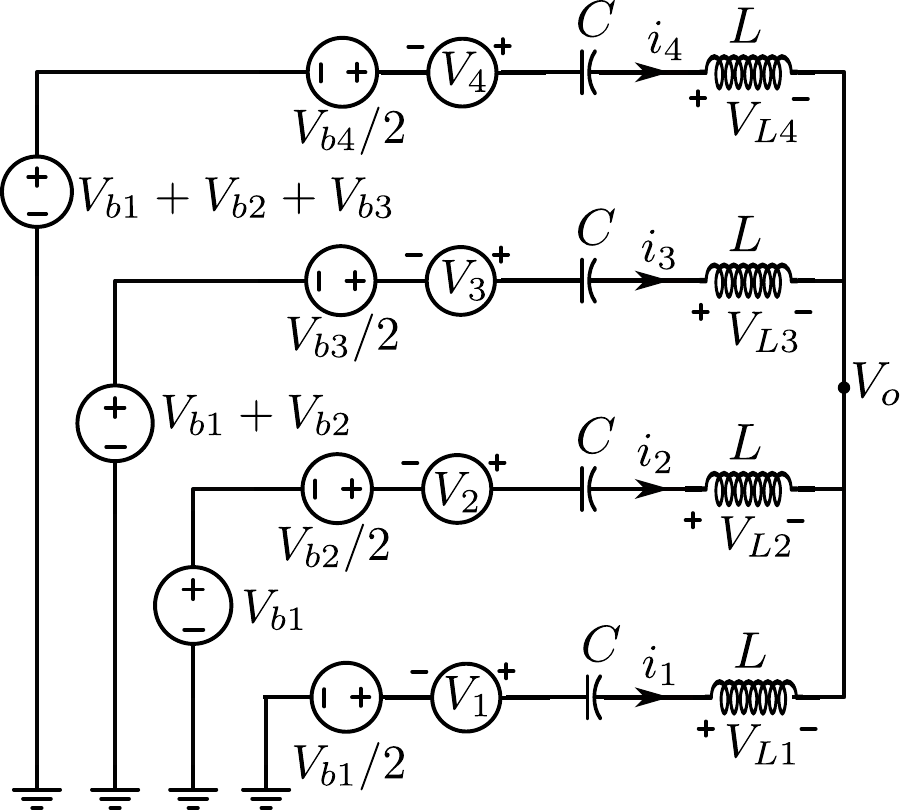}} \end{subfigure}	\hspace{0.2cm}
			\begin{subfigure}[]{\includegraphics[width=3.3cm]{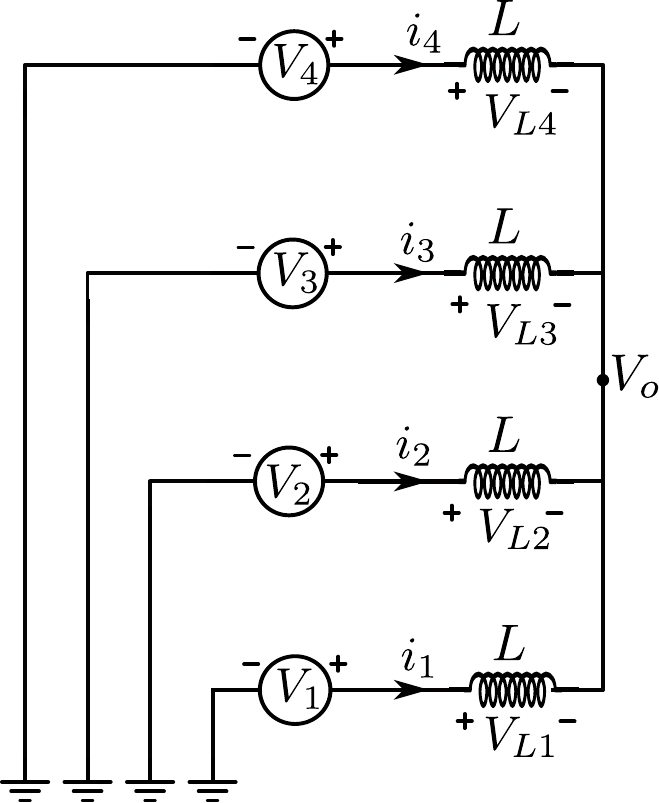}} \end{subfigure}
			\caption{(a) Equivalent circuit of the equalizer circuit for four batteries, (b) ac equivalent circuit of (a).}
			\label{eqv4}
		\end{figure}

\begin{figure}[h!]
		\centering
		\includegraphics[width=6cm]{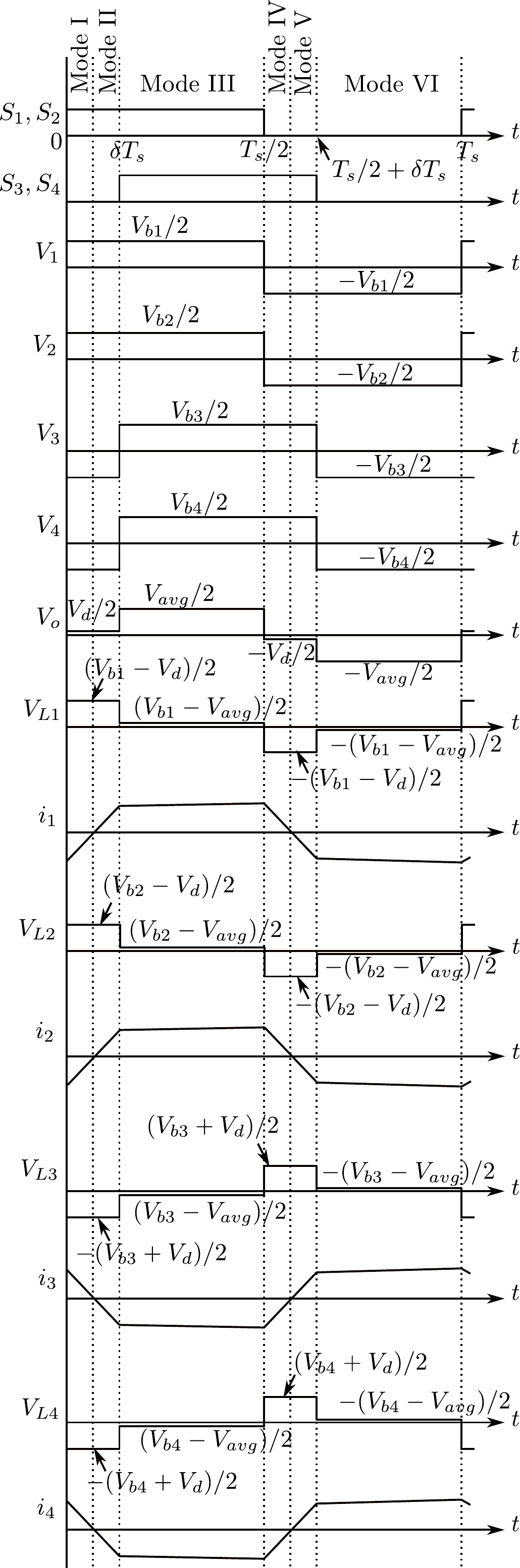}
		\caption{Voltage and current waveforms in four cell equalizer circuit in different modes of operations.}
		\label{wf4}
	\end{figure}

	A equivalent circuit is obtained for four batteries in Fig.\,\ref{eqv4}(a) using (\ref{vpk1}). Since, the dc blocking capacitor $C$ blocks the dc voltages in the circuit in Fig.\,\ref{eqv4}(a) and does not offer significant impedance at switching frequency, the ac equivalent circuit is derived as shown in Fig.\,\ref{eqv4}(b), where,
	\begin{eqnarray}
	\label{vk}
		V_k=\frac{V_{bk}}{2}Sq(t+\delta_k T_s)
	\end{eqnarray}
	Since, battery 1 and 2 discharging and battery 3 and 4 charging, $\delta_k=0$ for $k=1,2$ and  $\delta_k=-\delta$ for $k=3,4$. The gate signals to the top devices are given by,
	\begin{eqnarray}
	\label{s12}
		&&S_1=S_2=\frac{1}{2}+\frac{1}{2}Sq(t)\\
		\label{s34}
		&&S_3=S_4=\frac{1}{2}+\frac{1}{2}Sq(t-\delta T_s)
	\end{eqnarray}
	The voltages $V_1$, $V_2$, $V_3$ and $V_4$ in Fig.\,\ref{eqv4}(b) are plotted in Fig.\,\ref{wf4} along with the gate signals in (\ref{s12}) and (\ref{s34}). Using superposition theorem in Fig.\ref{eqv4}(b), it can be shown that,
	\begin{eqnarray}
	\label{vo}
		V_o=\frac{V_1+V_2+V_3+V_4}{4}
	\end{eqnarray}
	$V_o$ is plotted in Fig.\,\ref{wf4} using (\ref{vo}).  Voltages across the inductors can be derived from plots of $V_1,V_2,V_3,V_4$ and $V_o$ as shown in Fig.\,\ref{wf4}. The voltages across the inductors are used to obtain the waveforms of the inductor currents. These inductor currents are used to explain the operation of the circuit in different modes.

	\begin{figure}[h!]
		\centering
		\begin{subfigure}[]{\includegraphics[width=4.2cm]{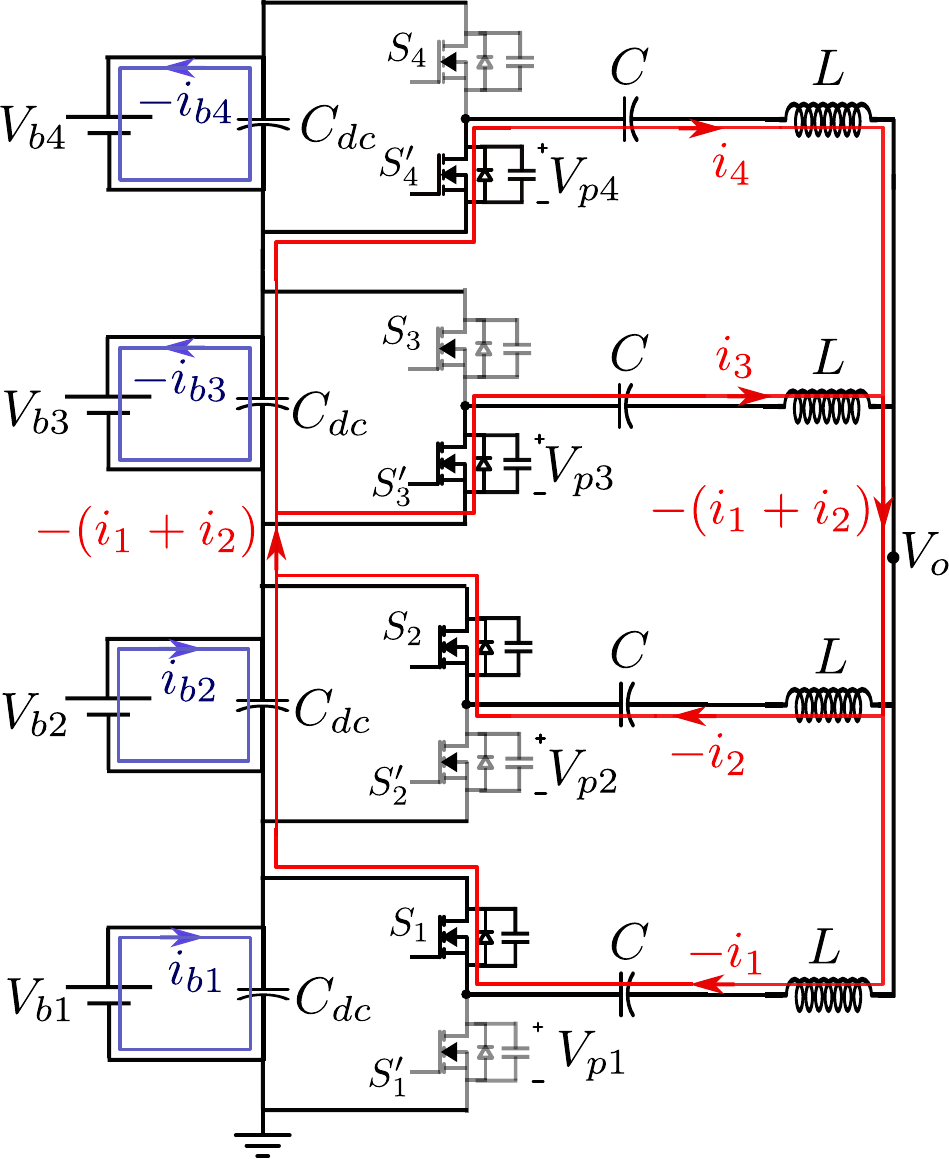}} \end{subfigure}	\hspace{0.0cm}
		\begin{subfigure}[]{\includegraphics[width=4.2cm]{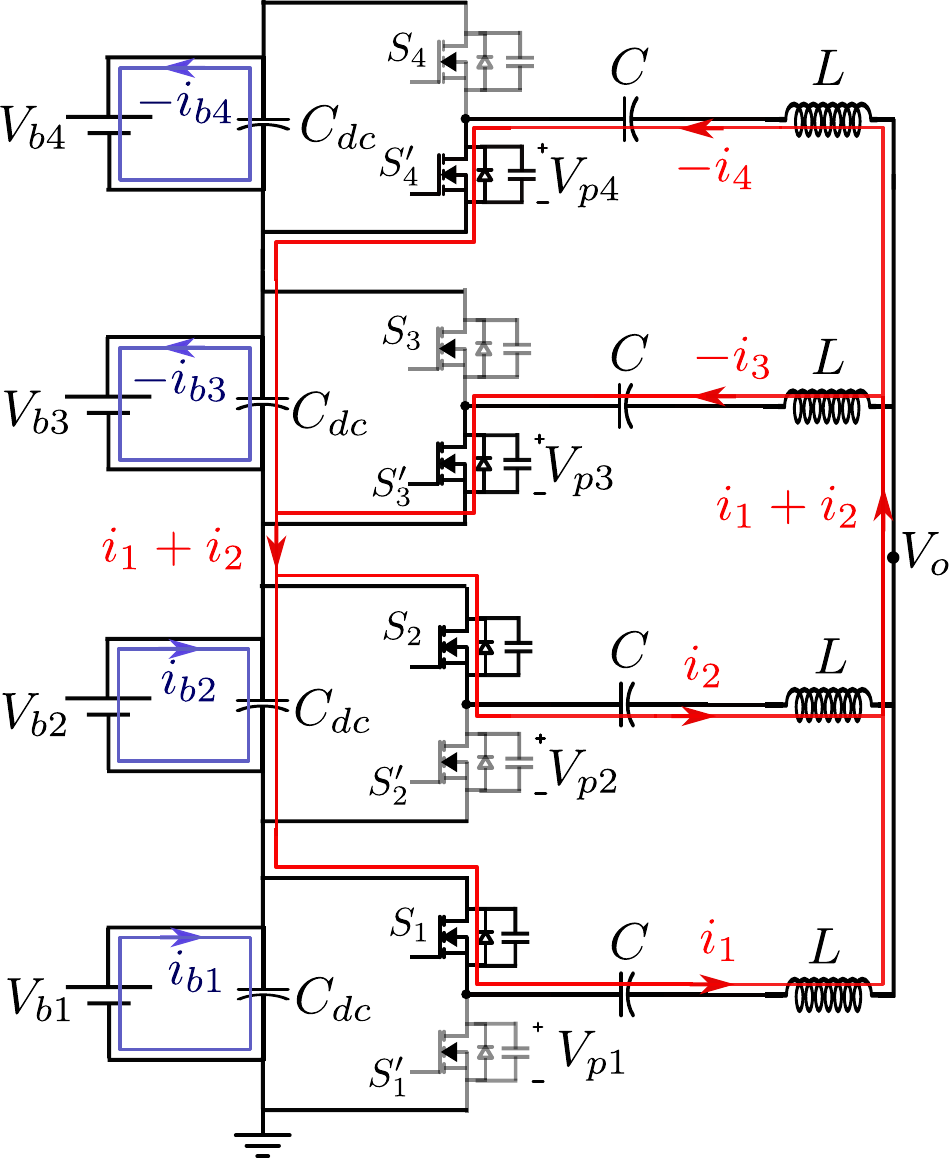}} \end{subfigure}
		
		\begin{subfigure}[]{\includegraphics[width=4.2cm]{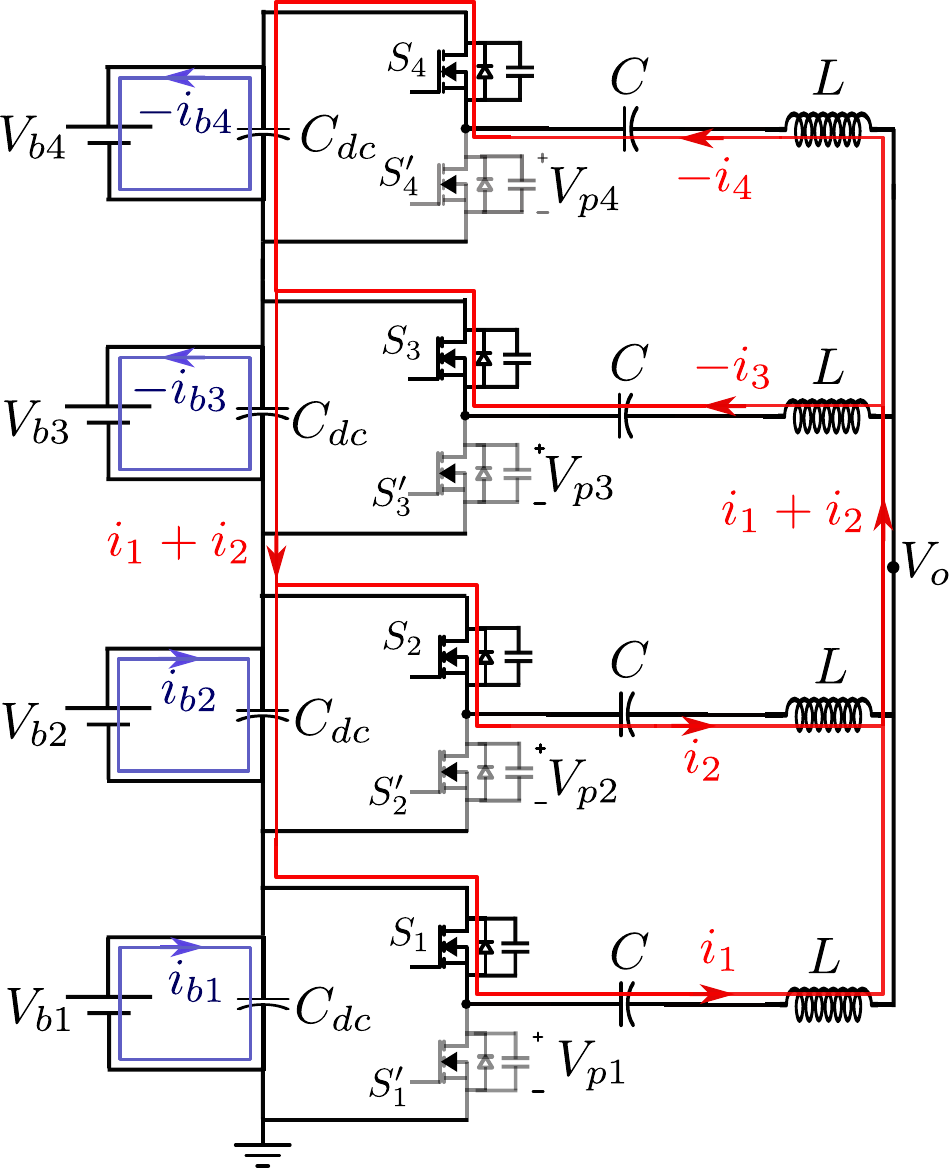}} \end{subfigure}	\hspace{0.0cm}
		\begin{subfigure}[]{\includegraphics[width=4.2cm]{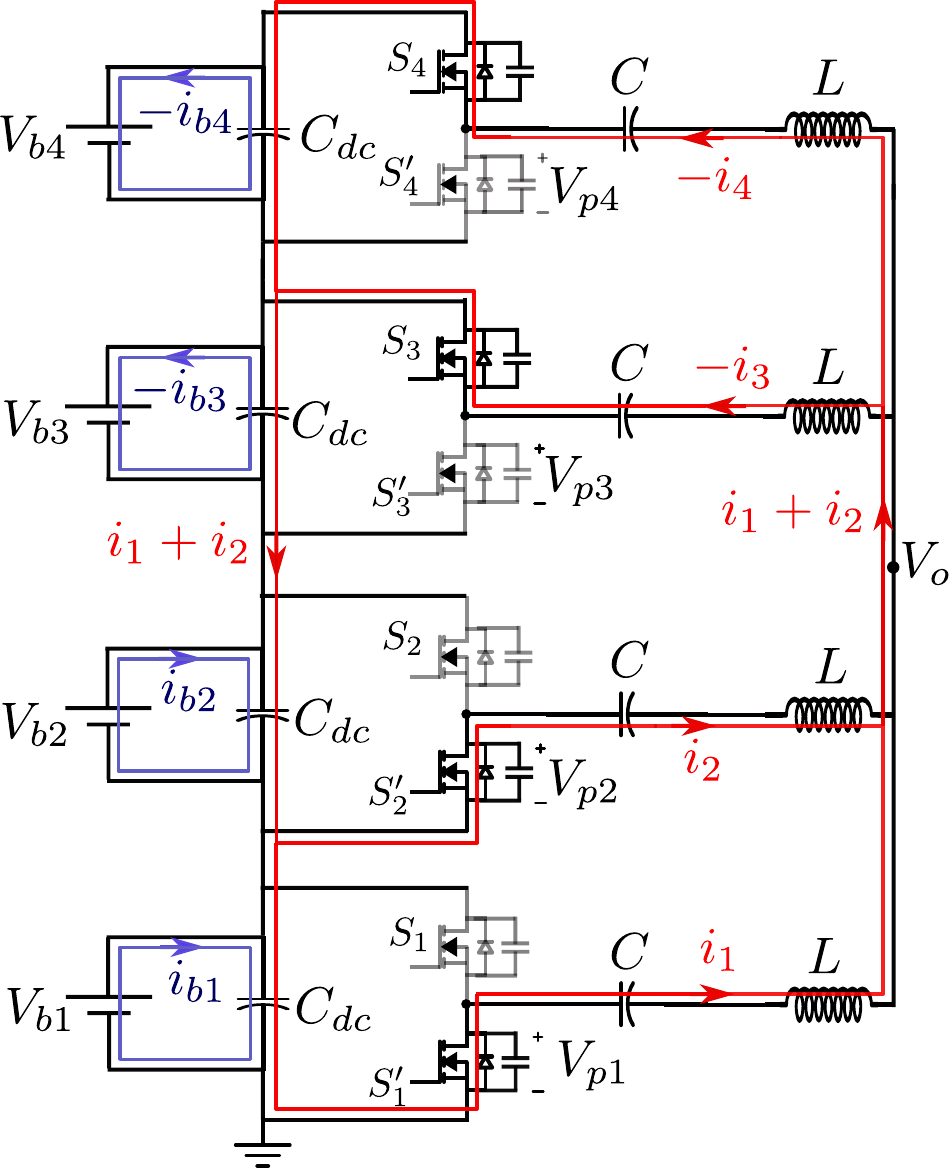}} \end{subfigure}
		
		\begin{subfigure}[]{\includegraphics[width=4.2cm]{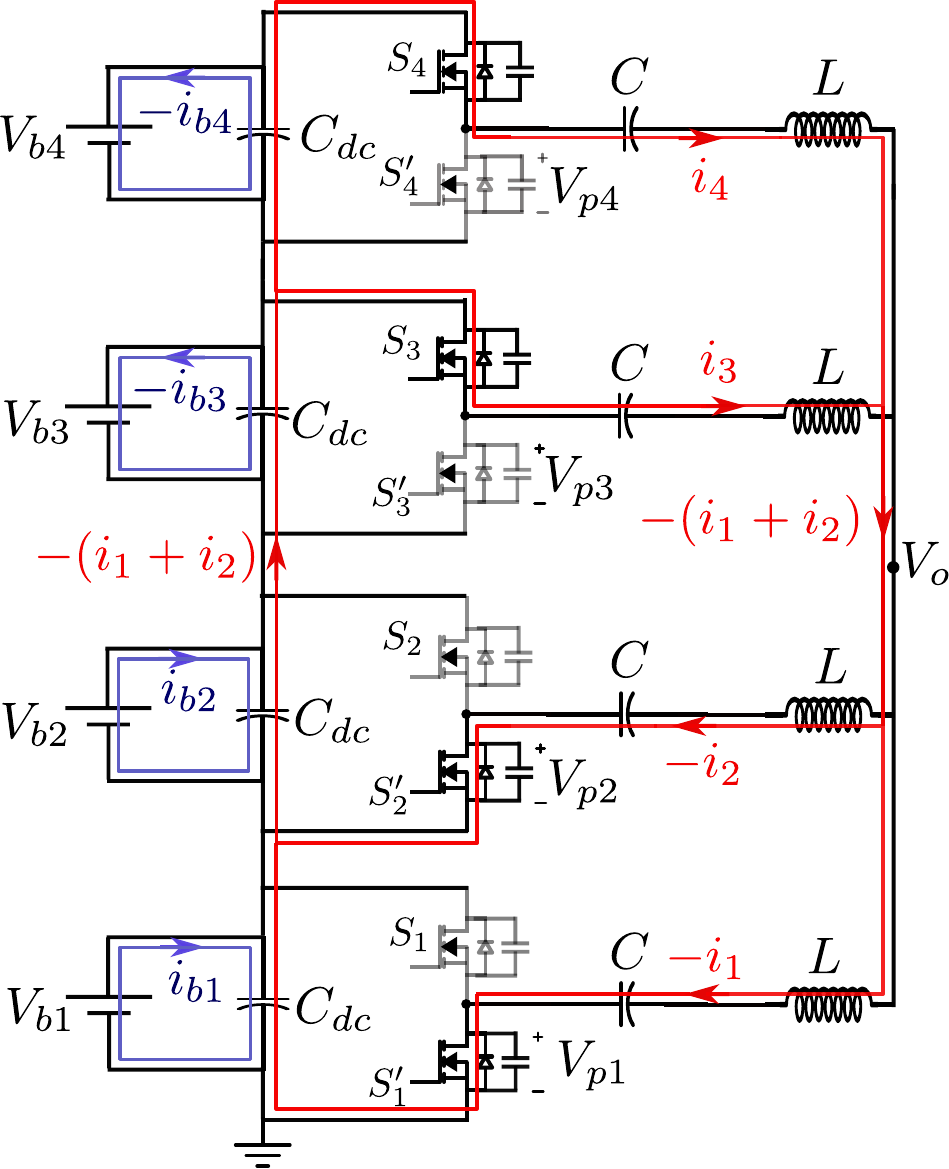}} \end{subfigure}	\hspace{0.0cm}
		\begin{subfigure}[]{\includegraphics[width=4.2cm]{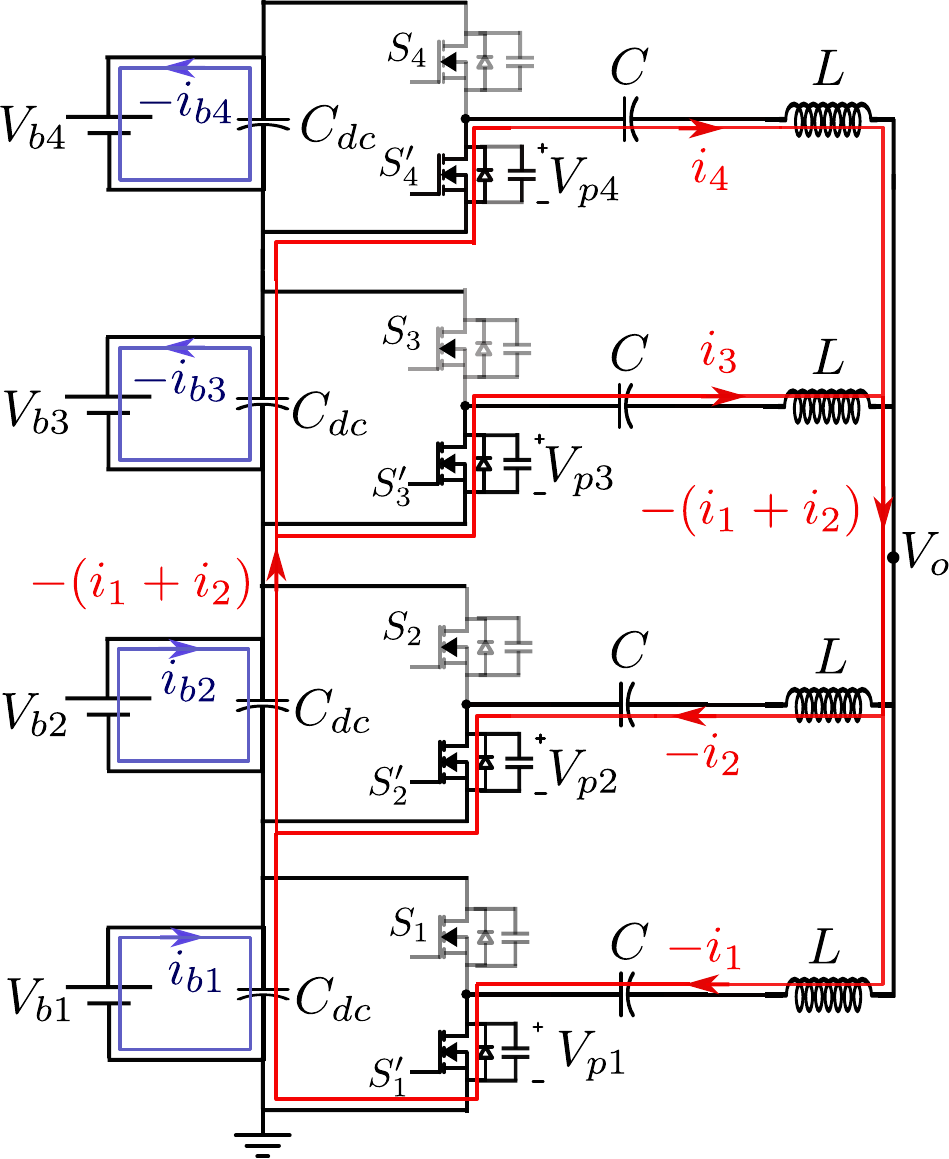}} \end{subfigure}
		\caption{Different modes of operations of a four cell equalizer, (a) mode I, (b) mode II, (c) mode III, (d) mode IV, (e) mode V and (f) mode VI.}
		\label{modes}
	\end{figure}	
		
	The circuit operation has six modes of operation as shown in Fig.\,\ref{wf4}. Each mode is shown with the current flow path, direction and current value. The effect of dead time is not considered here.	
	\begin{itemize}
		\item \textit{Mode I and II}: In mode I and II top devices of converter 1 and 2 and bottom devices of converter 3 and 4 are conducting. Current flow paths in both the modes are same, but the direction of current flow in one mode is opposite to other. If the slope of current in mode III is very small, then due to the symmetry of currents and duration of  modes I and II in Fig.\,\ref{wf4}, current flowing through the dc bus capacitors in these modes will only lead to a switching frequency ac current in them. However, if the slope of current in mode III is not negligible, then the net effect of the currents in mode I and II is to discharge $C_{dc}$ of second converter and charge $C_{dc}$ of third converter with a small amount of charge. Thus,  some power transfer takes place in these modes which is small compared to the power transfer in mode III and VI. So, mode III and VI decide which battery is charging and which battery is discharging.
		
		\item \textit{Mode IV and V}: In these modes, bottom devices of converter 1 and 2 and top devices of converter 3 and 4 are conducting. Similar to mode I and II, the current path is same in both the modes, but current direction in mode IV is opposite to the current direction in mode V. Thus, similar to mode I and II, the current flow in these modes mainly leads to a switching frequency ac current flowing through the dc bus capacitors and a small amount of power transfer happens which is small compared to mode III and VI.
		
		\item\textit{Mode III and VI}: In mode III, all the top switches of all four converters are conducting and mode IV, all the bottom switches are conducting. The fourth converter charges its dc bus capacitor in mode III, but does not charge or discharge it in mode VI. Thus, the charging current has a dc component which charges the battery 4. Similarly, the first converter does not charge or discharge its dc bus in mode III, but discharges it in mode VI. The dc component of the discharge current is used to discharge battery 1. Hence, charging of battery 4 and discharging of battery 1 are achieved in these two modes.\\
		
		Dc bus capacitor in third converter is charged with current $(i_3+i_4)$ in mode III and discharged with $i_4$ in mode VI. Due to equal duration of these two modes and symmetry of waveform of $i_4$, the current $i_4$ only produces a switching frequency ac current in the dc bus capacitor. Thus, the current $i_3$ in mode III effectively provides charging current to the dc bus capacitor and the dc component of this charging current charges battery 3. Similarly, the dc bus capacitor of the second converter is charged with current $i_1$ in mode III and discharged with current $(i_1+i_2)$ in mode VI. Due to equal duration of these two modes and symmetry of waveform of $i_1$ in Fig.\,\ref{wf4}, the current $i_1$ produces a switching frequency ac current in the dc bus capacitor and $i_2$ provides discharge current to the dc bus in mode VI. The dc component of this current leads to discharge of battery 2. Hence, battery 2 is discharging while battery 3 is charged.

	\end{itemize}
	
	Thus, power transfer from multiple batteries to multiple batteries is possible simultaneously with this topology.

	\subsection{Power and Current Calculation}%%%%%%%%%%%%%%%%%%%%%%%%%%%%%%%%%%%%%%%%%%%%%%%%%%%%%%%%%%%%%%%%%%%%%%%%%%%%%%%%%%%%%%%%%%%%%%%%%%%%%%%%%%%%%%%%%%%%%%%%%%%%%%%%%%%%%%%%%5
	\subsubsection{Inductor Current}
	The voltage and current waveforms are similar to a phase-shifted half-bridge converter. However, the expression for inductor current in this converter need to extended for $n$ number of converters connected together in the case of the proposed topology.
	The derivation of inductor current is provided in Appendix \ref{ind_current}.	The inductor current in the $k^{th}$ converter can be expressed as, 
	\begin{eqnarray}
	\label{ilk}
	i_{k}=\frac{T_s}{8nL}\left[nV_{bk}Tr(t+\delta_kT_s)-\sum_{i=1}^{n}V_{bi}Tr(t+\delta_iT_s)\right]
	\end{eqnarray}
	
	\subsubsection{Power Transfer}
	%Lets consider the functions $Sq(t+\delta_iT_s)$ and $Tr(t+\delta_jT_s)$ where $\delta_j>\delta_i$ and their product as shown in Fig.\,\ref{power}. 
	%Area under product waveform in one half cycle is given by,
	%\begin{eqnarray}
	%A_{S_iT_j}=2.\frac{1}{2}(\delta_j-\delta_i)T_s\left(1+1-\frac{4}{T_s}(\delta_j-\delta_i)T_s\right)
	%\end{eqnarray}
	%Now, using this, the following expression can be evaluated,
	%\begin{eqnarray}
	%\frac{1}{T_s}\int_{0}^{T_s}Sq(t+\delta_iT_s)Tr(t+\delta_jT_s)dt=\frac{2}{T_s}A_{S_iT_j}&& \nonumber\\
	%=4(\delta_j-\delta_i)\left(1-2(\delta_j-\delta_i)\right)&&
	%\end{eqnarray}
	%
	%\begin{figure}[h!]
	%	\centering
	%	\includegraphics[width=8.5cm]{Drawings/power.pdf}
	%	\caption{(a) Waveforms of functions $Sq(t+\delta_iT_s)$ and $Tr(t+\delta_jT_sx)$, (b) waveform of product of these two functions.}
	%	\label{power}
	%\end{figure}
	%
	%
	%When $\delta_j<\delta_i$, following the same procedure, the above integral can be evaluated as follows,
	%\begin{eqnarray}
	%\frac{1}{T_s}\int_{0}^{T_s}Sq(t+\delta_iT_s)Tr(t+\delta_jT_s)dt=\frac{2}{T_s}A_{S_iT_j}&&\nonumber\\
	%=-4(\delta_i-\delta_j)\left(1-2(\delta_i-\delta_j)\right)&&
	%\end{eqnarray}
	%
	%Hence in general,
	%\begin{eqnarray}
	%\label{avg_fn}
	%\frac{1}{T_s}\int_{0}^{T_s}Sq(t+\delta_iT_s)Tr(t+\delta_jT_s)dt=4(\delta_j-\delta_i)\left(1-2|(\delta_j-\delta_i)|\right)
	%\end{eqnarray}
	
	Power transfer from $k^{th}$ source is given by,
	\begin{eqnarray}
	P_k &=& \frac{1}{T_s}\int_{0}^{T_s}V_{pk} i_{k}dt\\
	&=& \frac{1}{T_s}\int_{0}^{T_s}\left[\frac{V_{bk}}{2}+\frac{V_{bk}}{2}Sq(t+\delta_k T_s)\right] i_{k}dt
	\end{eqnarray}
	Since, the current $i_k$ has only switching frequency component and its harmonics,
	\begin{eqnarray}	
	P_k &=& \frac{1}{T_s}\int_{0}^{T_s}\left[\frac{V_{bk}}{2}Sq(t+\delta_k T_s)\right] i_{k}dt\\
	&=& \frac{1}{T_s}\int_{0}^{T_s}\frac{V_{bk}}{2}Sq(t+\delta_kT_s)
	\frac{T_s}{8nL}\bigg[nV_{bk}Tr(t+\delta_kT_s)\nonumber\\
	\label{pk_int}
	&&-\sum_{i=1}^{n}V_{bi}Tr(t+\delta_iT_s)\bigg]dt
	%
	%&=& \frac{1}{T_s}\int_{0}^{T_s}\frac{V_{bk}T_s}{16nL}Sq(t+\delta_kT_s)
	%\left[nV_{bk}Tr(t+\delta_kT_s)-\sum_{i=1}^{n}V_{bi}Tr(t+\delta_iT_s)\right]dt
	\end{eqnarray}
%	The following integral can be evaluated,
%	\begin{eqnarray}
%	&&\frac{1}{T_s}\int_{0}^{T_s}Sq(t+\delta_iT_s)Tr(t+\delta_jT_s)dt \nonumber\\
%	&&=
%	\begin{cases}
%	4(\delta_j-\delta_i)\left(1-2(\delta_j-\delta_i)\right), \text{    when }\delta_j>\delta_i\\
%	-4(\delta_i-\delta_j)\left(1-2(\delta_i-\delta_j)\right), \text{    when }\delta_j<\delta_i
%	\end{cases}\\
%	\label{avg_fn}
%	&&=4(\delta_j-\delta_i)\left(1-2|(\delta_j-\delta_i)|\right)
%	\end{eqnarray}

	The expression for power is obtained by evaluating the integral in (\ref{pk_int}) and can be expressed as follows,	
	\begin{eqnarray}
	\label{pk}
	P_k&=& \frac{V_{bk}}{4nLf_s}
	\left[\sum_{i=1}^{n}V_{bi}(\delta_k-\delta_i)\left(1-2|(\delta_k-\delta_i)|\right)\right]
	\end{eqnarray}	
	It can be observed from (\ref{pk}) that the power transfer level between batteries does not depend on the difference between the battery voltages.

	\subsubsection{Battery Current}
	Using KCL in Fig.\,\ref{schematic}, following expression can be obtained,
	\begin{eqnarray}
	i_{bk}+i_{Ck}=S_k i_k-\sum_{j=1}^{k}i_j
	\end{eqnarray}	
	Here, $S_k$ is the gate signal of the top device of the $k^{th}$ converter. The sum of battery and dc bus capacitor currents has a dc component and switching frequency components. Most of the ac component flows through the capacitor, $C_{dc}$. DC component in battery current is given by,
	\begin{flalign}
	\label{ibk}
	I_{bk}&=\frac{P_k}{V_{bk}}\nonumber\\
	&=\frac{1}{4nLf_s}\left[\sum_{i=1}^{n}V_{bi}(\delta_k-\delta_i)\left(1-2|(\delta_k-\delta_i)|\right)\right]
	\end{flalign}
	
	The expression of battery current in (\ref{ibk}) does not depend on the voltage differences of the batteries. Thus the equalization power does not reduce as the battery voltages come closer to each other with progress in the equalization process and the equalization process does not slow down with time.

%	\subsection{Voltage Balance Algorithm}%--------------------------------------------------------------------------------------------
%	An algorithm for controlling the phases of the gate drive signals is required for achieving convergence of voltages. A simple algorithm is sufficient for the proposed topology. Charging and discharging is done in open loop control. Only the phases of the gate drive signal has to decided based on the battery voltage. 
%	
%	Ideally, the battery voltages should be same and equal to their average voltage. So, the algorithm tries to charge or discharge a battery to bring its voltage within a tolerance band around the average of all the voltages. The voltage tolerance, $V_{tol}$ decides the upper limit, $V_h$ and the lower limit, $V_l$ on the acceptable range of battery voltage as follows,
%	\begin{eqnarray}
%		V_h=V_{avg}+V_{tol}\\
%		V_l=V_{avg}-V_{tol}
%	\end{eqnarray} 
%	In general the gate drive signal for $k^{th}$ converter is, $S_k=1/2+Sq(t+\delta_k T_s)/2$. If the battery voltage is higher than $V_h$, then $\delta_k$ is set to $0$ and if it is lower than $V_l$, then $\delta_k$ is set to $-\delta$ which is a predefined design quantity. $\delta T_s$ is the phase difference between charging and discharging converters. The voltage balance algorithm is shown in a flowchart in Fig.\,\ref{flowchart}. 
%	
%	
%	\begin{figure}[h!]
%		\centering
%		\includegraphics[width=8cm]{Drawings/flowchart.pdf}
%		\caption{Flowchart of the algorithm for controlling charging and discharging the $k^{th}$ battery.}
%		\label{flowchart}
%	\end{figure}

	\subsection{Battery Voltage in Acceptable Range}
	The battery with voltage in the acceptable range should not take part in power transfer. Hence the algorithm in Fig.\,\ref{flowchart} turns off both the switches of the converter when battery voltage is within the upper and lower limits. But, it is possible that the diode in parallel with the mosfet can turn on and act as a rectifier to charge the battery. The following analysis is done to find out the condition for diode turn-on.
	
\begin{figure}[h!]
		\centering
		\includegraphics[width=8cm]{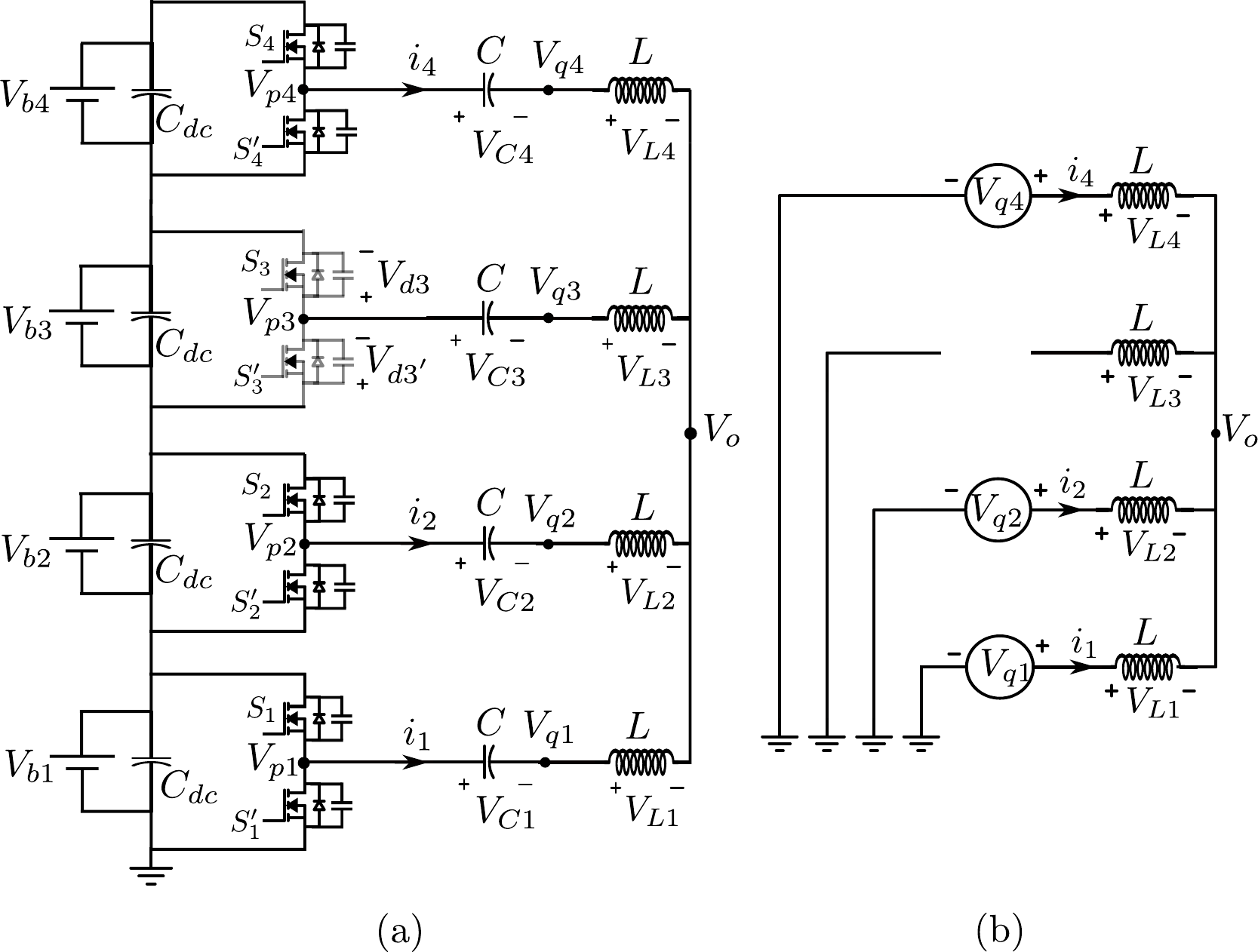}
		\caption{(a) Schematic diagram and (b) equivalent circuit of a four cell equalizer where one battery is not exchanging power.}
		\label{bat_in_range}
	\end{figure}

	Voltage equalizer circuit for four batteries are considered here as shown in Fig.\,\ref{bat_in_range}(a). The pole voltage of $k^{th}$ converter with respect to the negative terminal of $k^{th}$ battery is given in (\ref{vpk1}). The pole voltage of $k^{th}$ converter with respect to the ground is given by,
	\begin{eqnarray}
	V_{pk\_gnd}&=&\sum_{i=1}^{k-1}V_{bi}+V_{pk}\\
	\label{vpk_gnd}
	&=&\sum_{i=1}^{k-1}V_{bi}+\frac{V_{bk}}{2}+\frac{V_{bk}}{2}Sq(t+\delta_k T_s)
	\end{eqnarray}
	
	The voltage of the common point of all the inductors, $V_o$ has an switching frequency ac component $V_{o\_ac}$ and a dc component $V_{o\_dc}$,
	\begin{eqnarray}
	\label{vo1}
		V_o=V_{o\_ac}+V_{o\_dc}
	\end{eqnarray}
	Using KVL in Fig.\,\ref{bat_in_range} and using (\ref{vpk_gnd}) and (\ref{vo1}),
	\begin{eqnarray}
		&&V_{pk\_gnd}=V_{Ck}+V_{Lk}+V_o   \nonumber\\
		&&\Rightarrow\sum_{i=1}^{k-1}V_{bi}+\frac{V_{bk}}{2}+\frac{V_{bk}}{2}Sq(t+\delta_k T_s) \nonumber\\
		&&\qquad\qquad=V_{Ck}+V_{Lk}+V_{o\_ac}+V_{o\_dc}
	\end{eqnarray}
	Taking average on both sides over a switching period and rearranging terms, the voltage blocked by dc blocking capacitor is obtained,
	\begin{eqnarray}
	\label{vck}
	V_{Ck}=\sum_{i=1}^{k-1}V_{bi}+\frac{V_{bk}}{2}-V_{o\_dc}
	\end{eqnarray}
	
	The voltage of the common node of $C$ and $L$ with respect to the ground for $k^{th}$ converter is given by,
	\begin{eqnarray}
		V_{qk}&=&V_{pk\_gnd}-V_{Ck}\\
		&=& \frac{V_{bk}}{2}Sq(t+\delta_k T_s)+V_{o\_dc}
	\end{eqnarray}
	
	Now, it is assumed that the battery 3 is within the acceptable voltage range. Hence, the mosfets of the third converter are turned off and no current is flowing in the inductor as shown in Fig.\,\ref{bat_in_range}(a). So, the equivalent circuit in such condition is shown in Fig.\,\ref{bat_in_range}(b). Using superposition theorem, it can be shown that,
	\begin{eqnarray}
		V_o&=&\frac{V_{q1}+V_{q2}+V_{q3}}{3}\\
		\label{vo2}
		&=&V_{o\_dc}+\frac{1}{3}\sum_{k=1,k\ne 3}^{4}\frac{V_{bk}}{2}Sq(t+\delta_k T_s)
	\end{eqnarray}
	The voltage across the top diode of the third converter in Fig.\,\ref{bat_in_range}(a) is given by,
	\begin{eqnarray}
		V_{d3}=V_{C3}+V_o-\sum_{k=1}^{3}V_{bk}
	\end{eqnarray}
	Using (\ref{vck}) and (\ref{vo2}),
	\begin{eqnarray}
	\label{vd3}
	V_{d3}=\frac{1}{3}\sum_{k=1,k\ne 3}^{4}\frac{V_{bk}}{2}Sq(t+\delta_k T_s)-\frac{V_{b3}}{2}
	\end{eqnarray}
	Since, the positive peak of $Sq(t)$ function is $1$, the maximum possible forward voltage on the diode is,
	\begin{eqnarray}
	\label{vd3max}
	V_{d3(max)}=\frac{1}{3}\sum_{k=1,k\ne 3}^{4}\frac{V_{bk}}{2}-\frac{V_{b3(min)}}{2}
	\end{eqnarray}
	The voltage $V_{b3}$ is within the acceptable tolerance band ($V_{b3}\in[V_{avg}-V_{tol},V_{avg}+V_{tol}]$), so the minimum value of $V_{b3}$ is given by,
	\begin{eqnarray}
		V_{b3(min)}&=&V_{avg}-V_{tol}\\
		&=&\frac{V_{b1}+V_{b2}+V_{b3(min)}+V_{b4}}{4}-V_{tol}\\
		\label{vb3min}
		\Rightarrow V_{b3(min)}&=&\frac{V_{b1}+V_{b2}+V_{b3}}{3}-\frac{4}{3}V_{tol}
	\end{eqnarray}
	Using (\ref{vb3min}) in (\ref{vd3max}),
	\begin{eqnarray}
	V_{d3(max)}=\frac{2}{3}V_{tol}
	\end{eqnarray}
	The voltage across the bottom diode in the third converter is given by,
	\begin{eqnarray}
		V_{d3^{'}}&=&-V_{d3}-V_{b3}\\
		&=&-\frac{1}{3}\sum_{k=1,k\ne 3}^{4}\frac{V_{bk}}{2}Sq(t+\delta_k T_s)-\frac{V_{b3}}{2}
	\end{eqnarray}
	The negative peak of $Sq(t)$ function is $-1$. Hence, the maximum possible forward voltage across the bottom diode is given by,
	\begin{eqnarray}
	\label{vd3'max}
	V_{d3^{'}(max)}&=&\frac{1}{3}\sum_{k=1,k\ne 3}^{4}\frac{V_{bk}}{2}-\frac{V_{b3(min)}}{2}\\
	&=& \frac{2}{3}V_{tol}
	\end{eqnarray}
	
	So, both the diodes will not turn on if the following condition based on the forward voltage drop of the diode, $V_{d,on}$ can be ensured,
	\begin{eqnarray}
		V_{d,on}>\frac{2}{3}V_{tol}
	\end{eqnarray}
	 In this work, a voltage tolerance of $25mV$ is considered which is lower than the forward voltage drop of the diodes. Thus, it can be ensured that the battery with voltage in acceptable range is neither charged nor discharged.
	
	\begin{figure}[h!]
		\centering
		\includegraphics[width=6 cm]{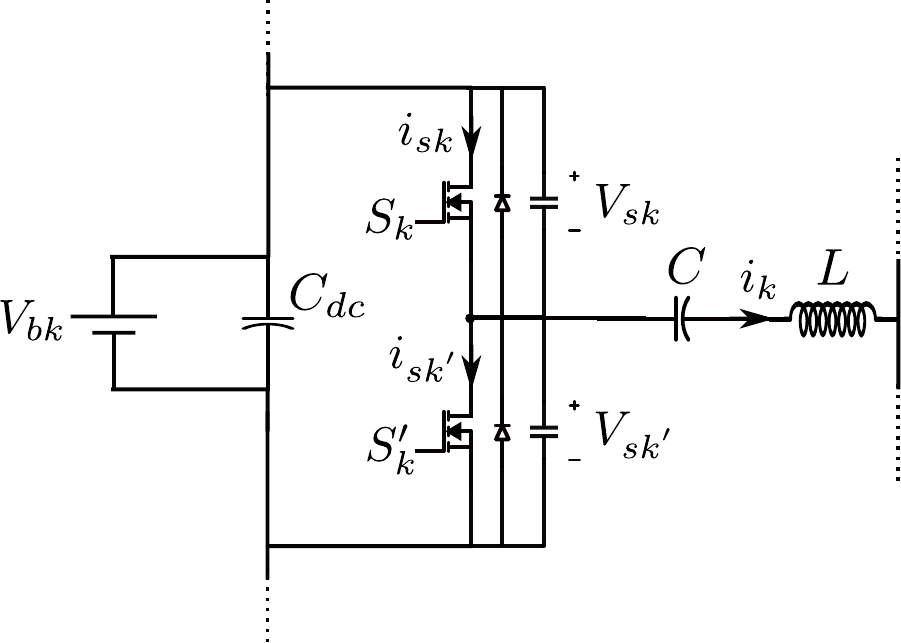}
		\caption{Schematic diagram of the $k^{th}$ converter.}
		\label{soft_sw_ckt}
	\end{figure}

	\section{Soft-switching}%%%%%%%%%%%%%%%%%%%%%%%%%%%%%%%%%%%%%%%%%%%%%%%%%%%%%%%%%%%%%%%%%%%%%%%%%%%%%%%%%%%%%%%%%%%%%%%%%%%%%%%%%%%%%%%%%%%%%%%%%%%%%%%%%%%%%%%%%5

		\begin{figure}[h!]
			\centering
			\includegraphics[width=6 cm]{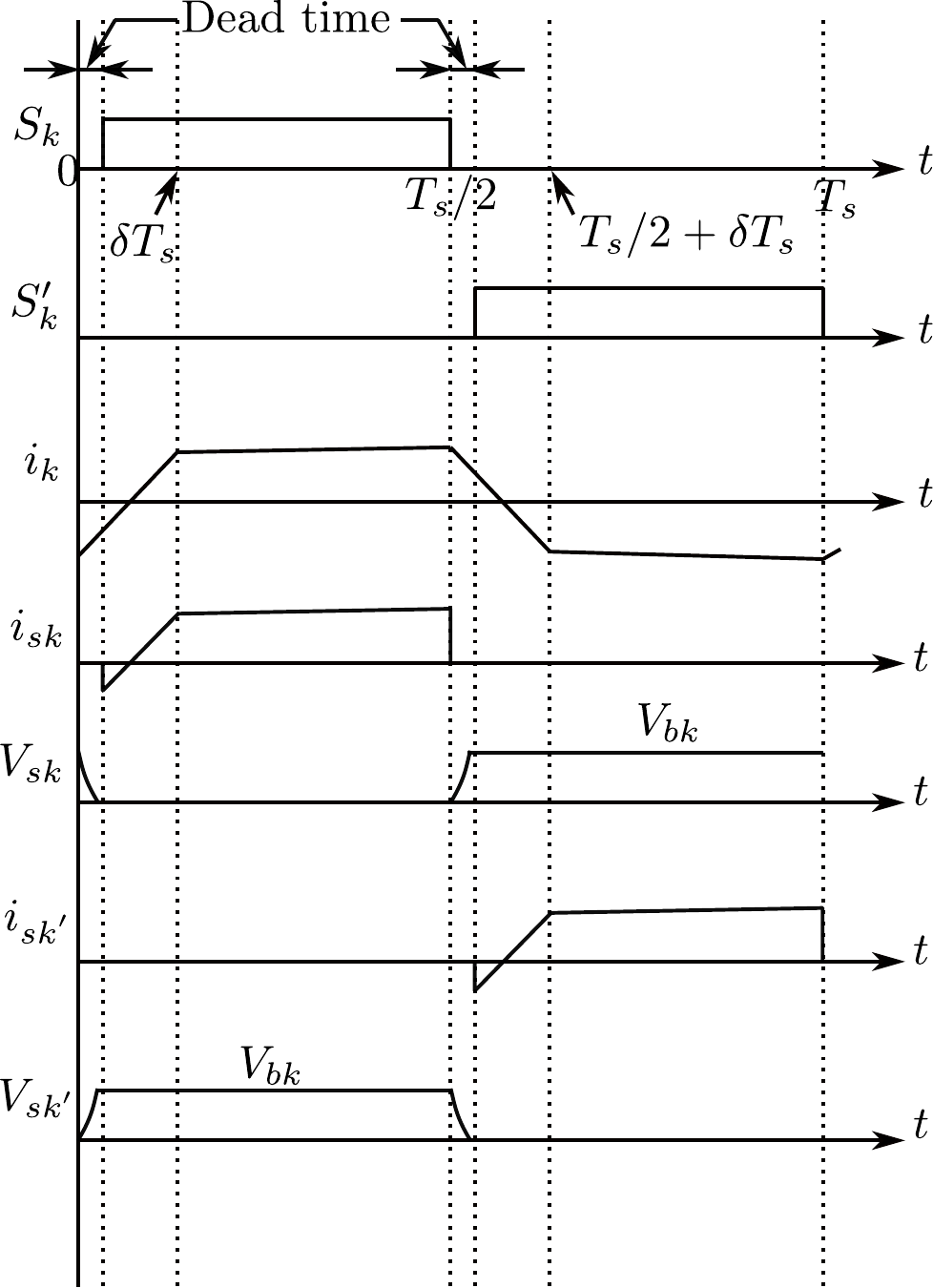}
			\caption{Voltages across and currents in the mosfets in $k^{th}$ converter. }
			\label{wf_soft}
		\end{figure}

	\subsection{Turn-on Transition}%--------------------------------------------------------------------------------------------------------------
	Fig.\,\ref{soft_sw_ckt} shows the $k^{th}$ converter where the $k^{th}$ battery is discharging. The current and voltage waveforms of $k^{th}$ converter are shown in Fig.\,\ref{wf_soft}. It can be observed from Fig.\,\ref{wf_soft}, the turn-on transition for the top device begins at $t=0$. However, the device turns on only after a small dead time. The inductor current is negative during this transition. Since both the devices are turned off during dead time, the negative current flows through the top diode and the top device turns on under zero voltage switching (ZVS) condition. The condition for ensuring ZVS is that inductor current is negative when the mosfet turns on. The inductor current is not controlled in close loop. However, it is possible to find a upper limit on inductor current at $t=0$ for all battery condition as shown below,	
	\begin{eqnarray}
	\label{imin}
		i_{k(t=0)}<-\frac{\delta}{2nLf_s}V_{bmin}
	\end{eqnarray}
	The derivation of this limit is provided in Appendix \ref{min_current}. 	
	Similarly, the turn-on transition for the bottom device begins at $t=T_s/2$. During the dead time, the inductor current is positive which has the following lower limit,
	\begin{eqnarray}
	i_{k(t=T_s/2)}>\frac{\delta}{2nLf_s}V_{bmin}
	\end{eqnarray}
	 Hence, the current flows through the bottom diode and the bottom device turns on under ZVS condition. Thus, ZVS turn on is ensured for all the devices of converters for discharging batteries under all load conditions. A similar argument can show that ZVS is ensured for converters for charging batteries as well.
	
	%\begin{figure}[h!]
	%	\centering
	%	\includegraphics[scale=0.7]{Drawings/schematic_soft_sw.pdf}
	%	\caption{Schematic diagram of the topology with modification for soft switching.}
	%	\label{schematic_soft}
	%\end{figure}
	
	\subsection{Turn-off Transition}%-------------------------------------------------------------------------------------------------------------
\begin{figure}[h!]
		\centering
		\includegraphics[width=6 cm]{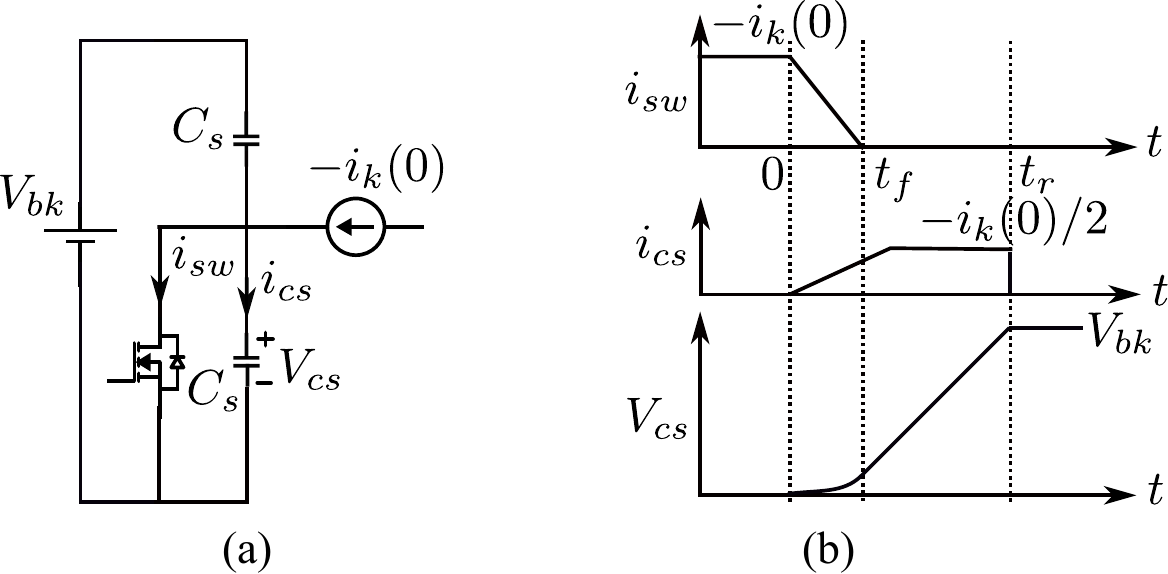}
		\caption{Current fall during turn off transition of bottom device, (a) equivalent circuit, (b) current and voltage waveforms. }
		\label{toff_fall}
	\end{figure}

	In order to achieve soft-switching during turn off transitions, an additional capacitor, $C_s$ is connected in parallel with each device as shown in Fig.\,\ref{schematic}. The energy stored in these capacitors during turn off transition is recycled during turn-on transition.
	The presence of $C_{s}$ capacitors reduces power loss during turn-off transition by slowing down the voltage rise across the switching device. The value of $C_s$ should be large enough so that the reduced power loss is small compared to a hard switched turn-off transition. During turn-off transition, inductor can be replaced by a current source as shown in Fig.\,\ref{toff_fall}(a) since the change in inductor current is negligible during current fall duration, $t_f$ and capacitor voltage rise duration $t_r$.

	The turn-off transition of the bottom device in Fig.\,\ref{toff_fall}(a)  starts at $t=0$ when the inductor current is $i_k(0)$ for $k^{th}$ converter. Current fall in the device is assumed to be linear. The rest of the inductor current is equally divided between the two capacitors. Since the $i_k$ is negative at $t=0$ as shown in Fig.\,\ref{wf_soft}, the bottom capacitor gets charged till its voltage reaches the battery voltage $V_{bk}$. The switch current, capacitor current, and the capacitor voltage during turn-off transition of the bottom device are shown in Fig.\,\ref{toff_fall}(b). The power loss during this turn-off transition is given by\cite{divan},
	\begin{eqnarray}
	%P_{toff\_soft}&=&\left[\int_{0}^{t_f} V_{cs}(t)i_{sw}(t)dt\right]f_s\nonumber\\
	P_{toff\_soft}&=&\frac{i_k(0)^2t_f^2f_s}{48C_s}
	\end{eqnarray}
	During hard switching, the voltage rise depends on the device. The voltage rise time $T_{vr}$ is a function of device characteristics and gate resistance.
	The power loss in a hard switched turn-off transition is given by\cite{eric},
	\begin{eqnarray}
	\label{p_toff_hard}
	P_{toff\_hard}=\frac{1}{2}V_{bk} (-i_k(0))(t_{vr}+t_f)
	\end{eqnarray}
	Hence, the ratio of turn-off losses in soft-switching and hard-switching is given by,
	\begin{eqnarray}
	\label{power_loss_ratio1}
	\frac{P_{toff\_soft}}{P_{toff\_hard}}=\frac{(-i_k(0))t_f^2}{24C_sV_b(t_{vr}+t_f)}	
	\end{eqnarray}

%	The ratio in (\ref{power_loss_ratio1}) is maximum in the worst case condition which happens if the inductor current at switching instance, $(-i_k(0))$ is maximum. Using (\ref{is3}), the worst case ratio of power losses is given by,
%	\begin{eqnarray}
%	\label{power_loss_ratio}
%	\frac{P_{toff\_soft}}{P_{toff\_hard}}=\frac{(-i_{k}(0))t_f^2}{24C_sV_{b(min)}(t_{vr}+t_f)}	
%	\end{eqnarray}
	This ratio is maximum when $(-i_k(0))$ is maximum. An upper limit on $(-i_k(0))$ can be found as follows,
	\begin{eqnarray}
	\label{imax}
		(-i_k(0)) \le (n-1)\frac{T_s}{8nL}[V_{bmax}-(1-4\delta)V_{bmin} )]
	\end{eqnarray}

\begin{figure}[h!]
		\centering
		\includegraphics[width=8.5 cm]{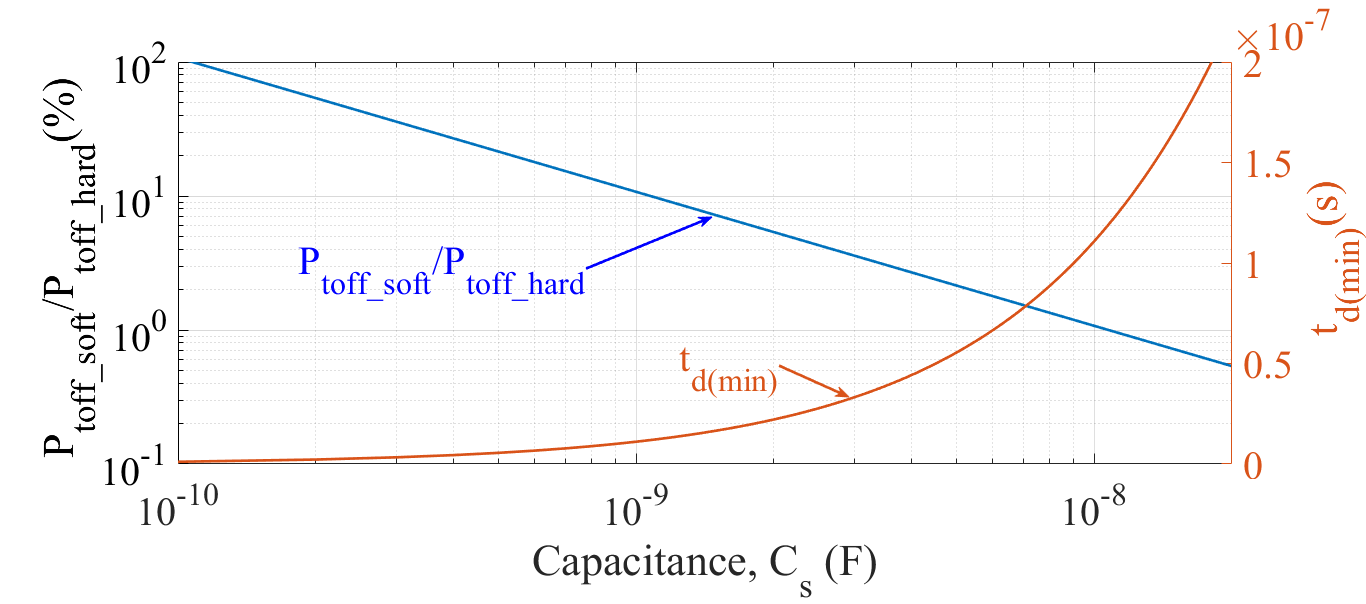}
		\caption{Plot of the ratio of turn-off power loss between soft-switched and hard-switched transitions and minimum required dead time with respect to the capacitance $C_s$.}
		\label{loss_cs}
	\end{figure}

	The derivation of this limit is provided in Appendix \ref{max_current}. The maximum value of $(-i_k(0))$ is calculated to be 13.6 A using (\ref{imax}) for $n=4$, $L=2.1\mu H$, $f_s=30 kHz$, $V_{bmax}=14.4 V$, $V_{bmin}=10.5V$ and $\delta=1/8$.  
	Voltage rise time $t_{vr}$ is calculated as $45.4 ns$ for battery voltage of 10.5 V and the current fall time $t_f$ for switching current of 13.6 A is calculated to be $10.6ns$  using gate resistance, $R_g=30\Omega$ and device characteristics from datasheet\cite{mosfet_data}. The ratio in (\ref{power_loss_ratio1}) is calculated for different values of $C_s$ and plotted in Fig.\,\ref{loss_cs}.

	Higher value of $C_s$ reduces power loss in MOSFET but increases required dead time. The required dead time must be more than capacitor voltage rise time $t_r$ in Fig.\,\ref{toff_fall}(b). It is assumed that $t_f<<t_r$ and inductor current does not change significantly in the duration $t_r$. Then the minimum required dead time $t_{d(min)}$ is given by,
	\begin{eqnarray}
	\label{tdmin}
	t_{d}\ge t_r=\frac{2C_sV_{bk}}{-(i_K(0))}
	\end{eqnarray}
	In the worst case, the minimum required dead-time is obtained for the minimum value of $(-i_k(0))$ which can be calculated from (\ref{imin}).	
	The minimum required dead time in (\ref{tdmin}) is plotted in Fig.\,\ref{loss_cs} along with the ratio of turn-off power losses in soft and hard switching. From Fig.\,\ref{loss_cs}, $C_s$ is chosen to be $4.7 nF$  as a trade-off between power loss and dead time requirement. However, the Mosfet has a parasitic capacitance between drain and source which varies from $1.2 nF$ to  $4.3 nF$. So, the effective value of $C_s$ varies from $5.9nF$ to $9nF$. Hence, from Fig.\,\ref{loss_cs},  the power loss in soft-switching is less than $1.5\%$ of that in hard-switching and the minimum required dead time is $120 ns$.
	
	From (\ref{p_toff_hard}), the hard switched turn-off loss for each mosfet is calculated to be $0.163 W$ for the worst case of maximum value of $(-i_k(0))$ and $V_{b}$. So, for eight mosfets the total hard-switched turn-off loss is $1.31 W$ which is $2.7\%$ of the rated power of the equalizer circuit. Thus, soft-switching helps to improve the efficiency by $2.7\%$ in the worst case. This improvement in efficiency will be greater if the circuit is designed for higher switching frequency. The switching frequency in the developed prototype is restricted to $30kHz$ mainly due to high conduction loss in the inductor windings at higher frequency. A better inductor design such as use of Litz wire in the winding can reduce the conduction loss leading to higher switching frequency. The proposed voltage equalizer can be implemented with lower cost and size with high efficiency with such inductor design.

	\begin{figure}[h!]		% hardware images
		\centering
		\begin{subfigure}[]{\includegraphics[width=5cm, height=3.7cm]{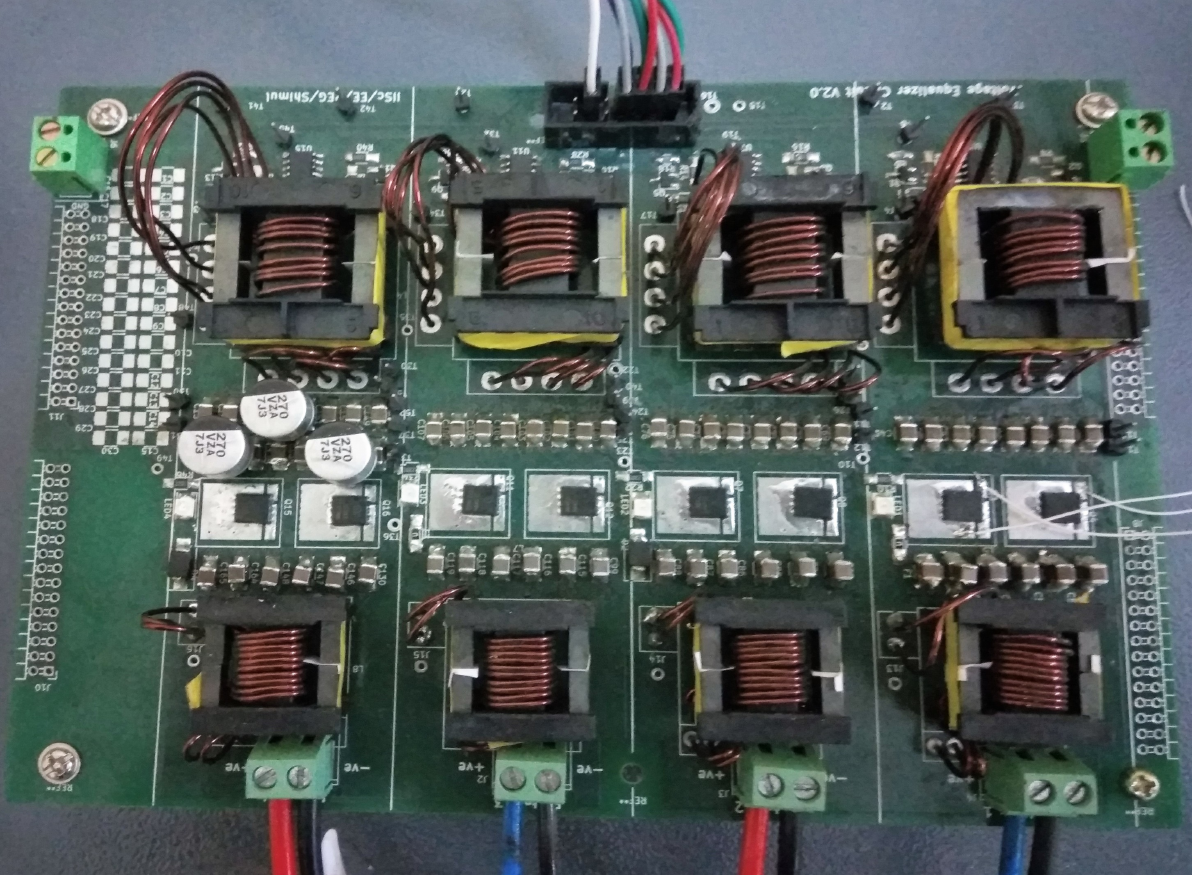}} \end{subfigure}	
		\begin{subfigure}[]{\includegraphics[width=3cm, height=3.7cm]{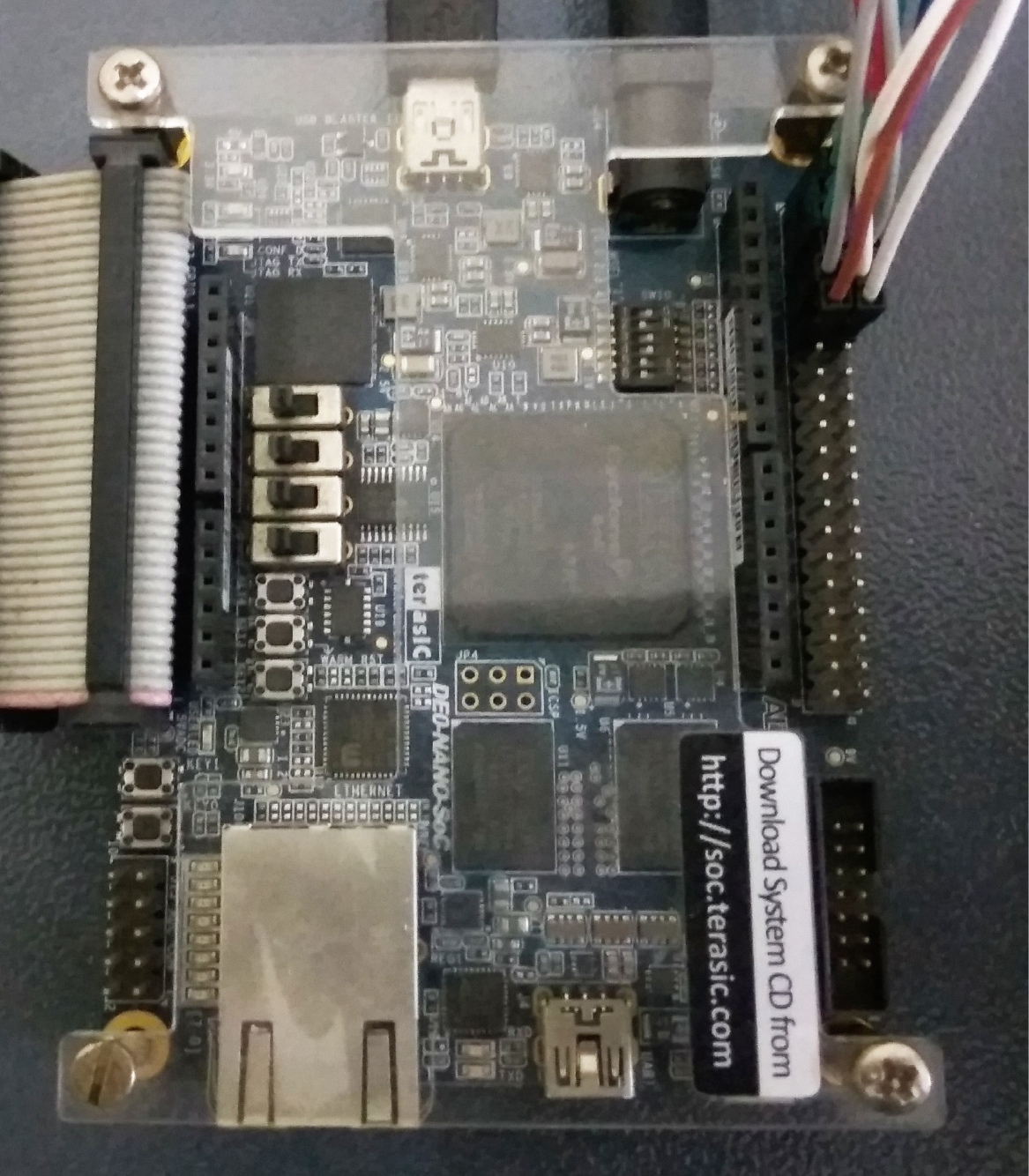}} \end{subfigure}
		\begin{subfigure}[]{\includegraphics[width=4.5cm,height=3.9cm]{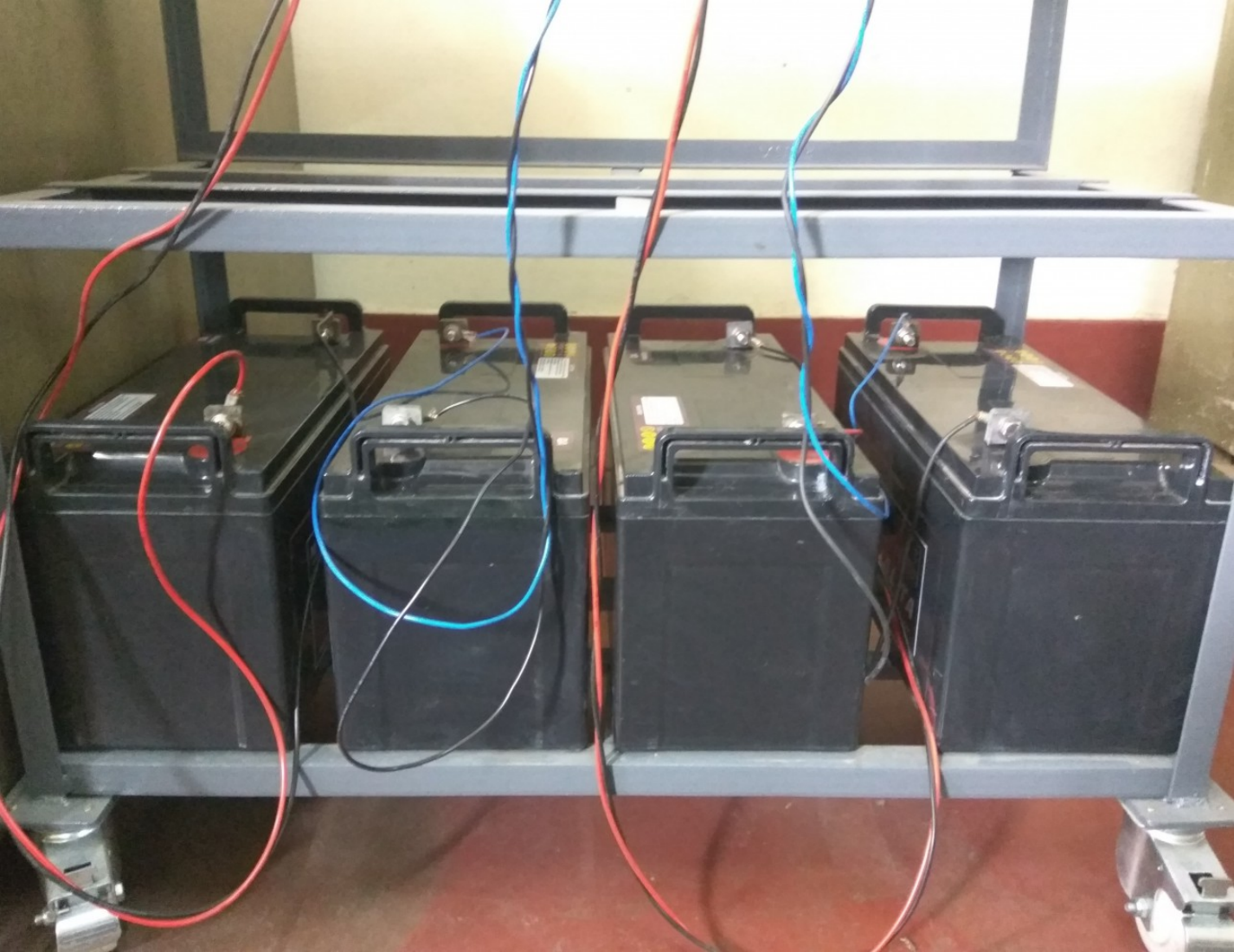}} \end{subfigure}	
		\begin{subfigure}[]{\includegraphics[width=3.5cm,height=3.9cm]{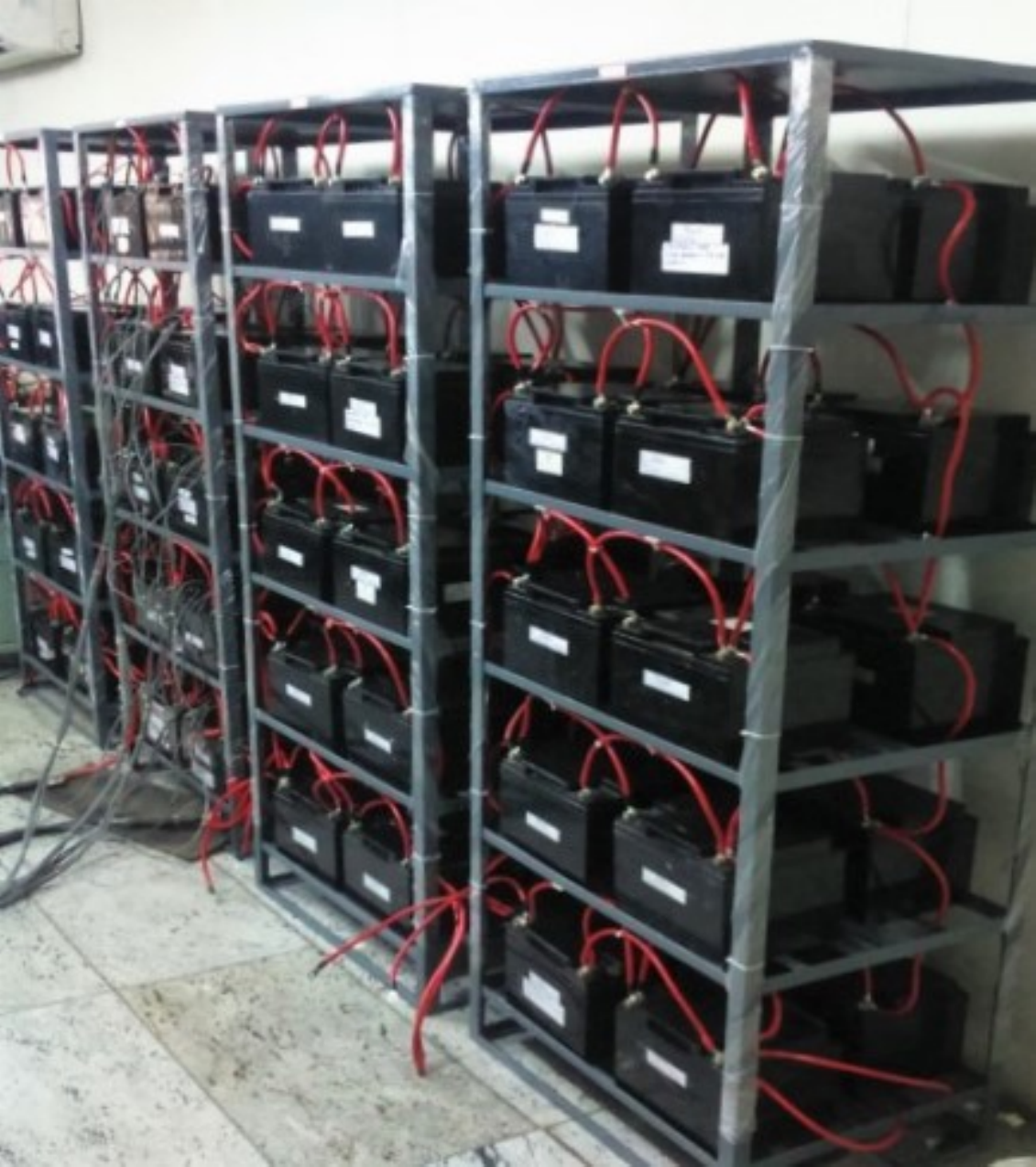}} \end{subfigure}	
		
		\caption{Hardware images: (a) voltage equalizer circuit, (b) FPGA based controller board, (c) battery bank, (d) hybrid ultra-capacitor (HUC) bank.}
		\label{hardware_image}
	\end{figure}

%%%%-----------------------------------------------------------------------------------------------------------------------------------------------------
\section{Topology Comparison}
A quantitative comparison of requirement of circuit components and efficiency as well as a qualitative comparison of equalization speed among some of the existing topologies  is presented in Table \ref{comp_tab}. It can be observed from this table that the proposed voltage equalizer circuit achieves all the desired properties of \cite{itec} and offers higher efficiency.

       \begin{table*}[t]
       	
       	\centering
       	\small
       	\caption{Comparison of the proposed voltage equalizer with some of the existing topologies.}
       	\renewcommand{\arraystretch}{1.5}
       	\renewcommand{\tabcolsep}{1pt}
       	\begin{tabular}{|c|c|c|c|c|c|c|c|c|c|c|c|}
       		\hline
       		\multirow{4}{*}{Topology} & \multirow{4}{*}{\parbox{3.8cm}{\centering{Type}}} & \multicolumn{4}{c|}{\multirow{2}{*}{Number of circuit elements}} & \multirow{4}{*}{\parbox{1.5cm}{\centering{Multi-cell to multi-cell charge transfer}}}    & \multirow{4}{*}{\parbox{1.5cm}{\centering{Ability to isolate a cell from equalization}}}  & \multirow{4}{*}{\parbox{1.7cm}{\centering{Decrease in battery current with voltage convergence }}} & \multirow{4}{*}{\parbox{1.2cm}{\centering{Soft-switching}}} & \multirow{4}{*}{\parbox{0.8cm}{\centering{Eff. (\%)}}} & \multirow{4}{*}{\parbox{1.6cm}{\centering{Equalization\\ speed }}} \\

       		&& \multicolumn{4}{c|}{} &  &  & & & &\\ \cline{3-6}

       		& & \multirow{2}{*}{Cap.} & \multirow{2}{*}{Ind.} & \multirow{2}{*}{Trans.} & \multirow{2}{*}{MOS.}  &  &  &  & & &\\ 
       		&&&&&&  &  &  & & &\\ \hline

       		Ref\cite{yua} & \parbox{3.8cm}{\vspace{0.1cm}\centering{Resonant Switched-capacitor based adjacent cell}\vspace{0.1cm}} & $n-1$ & $n-1$ & $0$ & $2n$ & No & No & Yes & Yes & 98.2 & Moderate\\ 
       		\hline
       		
       		Ref\cite{lim} & \parbox{3.8cm}{\vspace{0.1cm}\centering{Multi-winding transformer based cell to stack}\vspace{0.1cm}} & $0$ & $0$ & \parbox{1.0cm}{\centering{1 n-wind. }}  & $n$ & No & No & Yes & No & - & Satisfactory\\ 
       		\hline
       		
       		Ref\cite{lee} & \parbox{3.8cm}{\vspace{0.1cm}\centering{Selection switch based single-cell to single-cell}\vspace{0.1cm}} & $0$ & $0$ & $1$ & $2n+6$ & No & Yes & No & No  & 80.4 & Moderate\\ 
       		\hline
       		
       		Ref\cite{shang2} & \parbox{3.8cm}{\vspace{0.1cm}\centering{Switched-capacitor based multi-cell to multi-cell}\vspace{0.1cm}} & \parbox{0.6cm}{$\frac{n}{2}(n\\-1)$} & $0$ & $0$  & $2n$ & Yes & No & Yes  & No &  94.5 & Good\\ 
       		\hline
       		
       		Ref\cite{ye} & \parbox{3.8cm}{\vspace{0.1cm}\centering{Switched-capacitor based multi-cell to multi-cell}\vspace{0.1cm}} & $n$ & $0$ & $0$ & $2n$ & Yes & No & Yes  & No & - & Good\\ 
       		\hline
       		
       		Ref\cite{33} & \parbox{3.8cm}{\vspace{0.1cm}\centering{Isolated Cuk based multi-cell to multi-cell}\vspace{0.1cm}} & $3n$ & $4n$ & $0$ & $2n$ & Yes & Yes & No & No & 88.4 & Excellent\\ 
       		\hline
       		
       		\parbox{1.2cm}{\centering{Ref\cite{itec}}} & \parbox{3.8cm}{\centering{\vspace{0.1cm}Phase shifted half-bridge based multi-cell to multi-cell}\vspace{0.1cm}} & $n$ & $n$ & $0$ & $2n$ & Yes & Yes & No & No & 87.3 & Excellent\\ 
       		\hline
       		
       		\parbox{1.2cm}{\centering{Proposed topology}} & \parbox{3.8cm}{\centering{\vspace{0.1cm}Phase shifted half-bridge based multi-cell to multi-cell}\vspace{0.1cm}} & $n$ & $n$ & $0$ & $2n$ & Yes & Yes & No & Yes & 94.1 & Excellent\\ 
       		\hline
       		
       		\multicolumn{11}{l}{{$^*Note$: Cap.: Capacitor,  Ind.: Inductor, Trans.: Transformer, MOS.: MOSFET, Eff.: Efficiency, wind.: winding}}  
       		%              \multicolumn{9}{l}{  Err: error, G.D.: gate driver, Cur: current}
       	\end{tabular}
       	\label{comp_tab}
       	
       \end{table*}

	Apart from achieving higher efficiency, the proposed method of voltage equalization offers lower requirements of ratings of the circuit elements leading to a reduction in cost and size of the equalizer. In order to explain this improvement, required ratings of the passive elements for the equalizer in \cite{itec} and the proposed equalizer are compared in Table \ref{comp_ratings} for same power rating. 
	
	\begin{table}[h!]
		\centering
		\small
		\caption{A comparison between the topology in \cite{itec} and the proposed topology.}
		%    \resizebox{0.3\textwidth}{!}{
		\renewcommand{\arraystretch}{1.2}
		\renewcommand{\tabcolsep}{3pt}
		\begin{tabular}{|c|c|c|}
			\hline
			\parbox{3cm}{\centering{Circuit Parameters}} & \parbox{2cm}{\centering{Ratings for \cite{itec}}} & \parbox{2cm}{\centering{Ratings for proposed topology}}\\ \hline
			Nominal power & $12 W$ & $48 W$\\ \hline
			Switching frequency, $f_s$ & $50kHz$ & $30 kHz$ \\ \hline
			Inductor, $L$ & $30\mu H$ & $2.1\mu H$, $8A$ \\ \hline
			Dc blocking capacitor, $C$ & $940\mu F, 50V $ & $670 \mu F$, $50V$ \\ \hline
			Dc bus capacitor, $C_{dc}$ & $4.4mF, 35V$ & $640 \mu F$, $35V$ \\ \hline
			Capacitor for soft turn-off, $C_{s}$ & NA & $4.7n F$, $35V$ \\ \hline			
		\end{tabular}
		\label{comp_ratings}
	 \end{table}

\begin{table}[h!]
	\centering
	\small
	\caption{Design parameters and the passive components of the developed prototype.}
	%    \resizebox{0.3\textwidth}{!}{
	\renewcommand{\arraystretch}{1.2}
	\renewcommand{\tabcolsep}{6pt}
	\begin{tabular}{|c|c|}
		\hline
		Circuit Parameters & Ratings\\ \hline
		Nominal power & $48 W$\\ \hline
		Phase difference, ($\delta T_s$) & $T_s/8$\\ \hline
		Switching frequency, $f_s$ & $30 kHz$\\ \hline
		Inductor, $L$ & $2.1\mu H$, $8A$\\ \hline
		Dc blocking capacitor, $C$ & $670 \mu F$, $50V$\\ \hline
		Dc bus capacitor, $C_{dc}$ & $640 \mu F$, $35V$\\ \hline
		Capacitor for soft turn-off, $C_{s}$ & $4.7n F$, $35V$\\ \hline
		Power MOSFET & BSC009NE2LS5I, 25V, 100A\\ \hline

	\end{tabular}
	\label{ckt_par}
	%    }
	
\end{table}
	
	\section{Experimental Results}
	
	A 48 W prototype for four 12 V lead-acid batteries has been developed in the laboratory to verify the theoretically estimated circuit operation of the proposed soft-switched topology. An FPGA based digital controller board generates the gate pulses which control the charging and discharging of each battery. In order to test the performance of the developed equalizer, the following tests were performed,
	
	\begin{enumerate}
		\item Two battery are charged and two battery are discharged at the same time to verify multi-cell to multi-cell charge transfer capability of the equalizer as well as to measure the efficiency of the circuit.
		\item The equalizer is tested in a situation where one battery does not need be either charged or discharged to verify that this battery is not unnecessarily charged or discharged while the other batteries exchange power.
		\item The convergence of the battery voltages is tested to verify the effectiveness of the equalizer and the equalization algorithm.
		\item A hybrid ultra-capacitor (HUC) bank is chosen for testing the equalizer over multiple charge-discharge cycles as the HUCs take shorter time to complete one cycle compared to batteries. Also, this test verifies the effectiveness of the equalizer on an energy storage with higher charge-discharge rate. 
	\end{enumerate}
	
		\begin{figure}[h!]
			\centering
			\begin{subfigure}[]{\includegraphics[width=4cm,height=2.5cm]{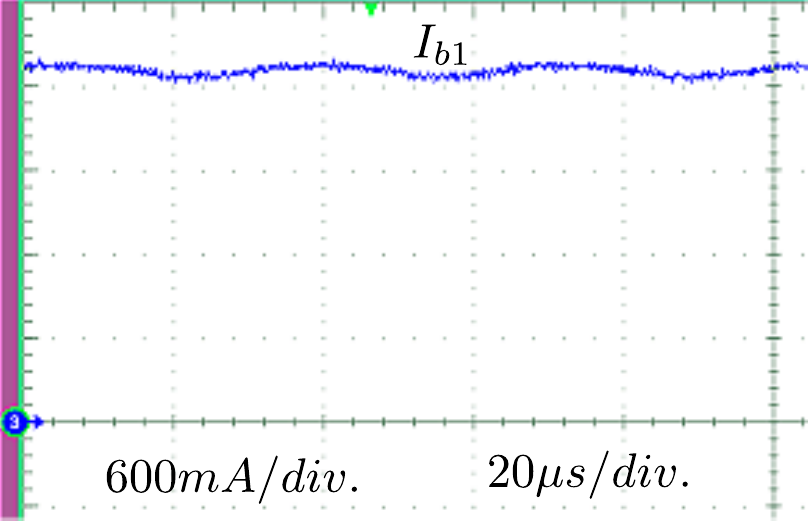}} \end{subfigure}	
			\begin{subfigure}[]{\includegraphics[width=4cm,height=2.5cm]{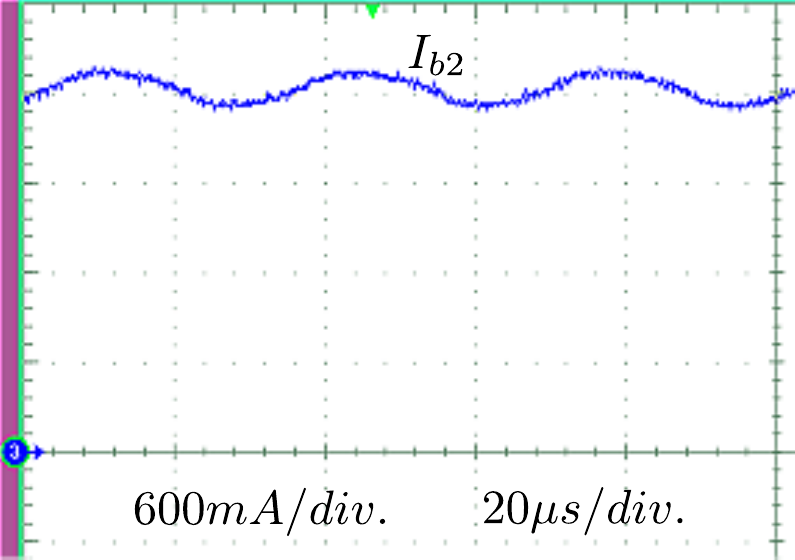}} \end{subfigure}
			\begin{subfigure}[]{\includegraphics[width=4cm,height=2.5cm]{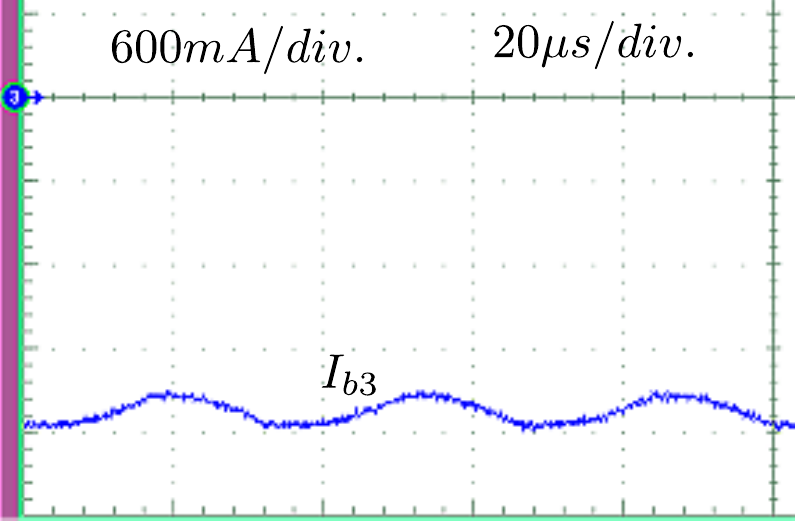}} \end{subfigure}	
			\begin{subfigure}[]{\includegraphics[width=4cm,height=2.5cm]{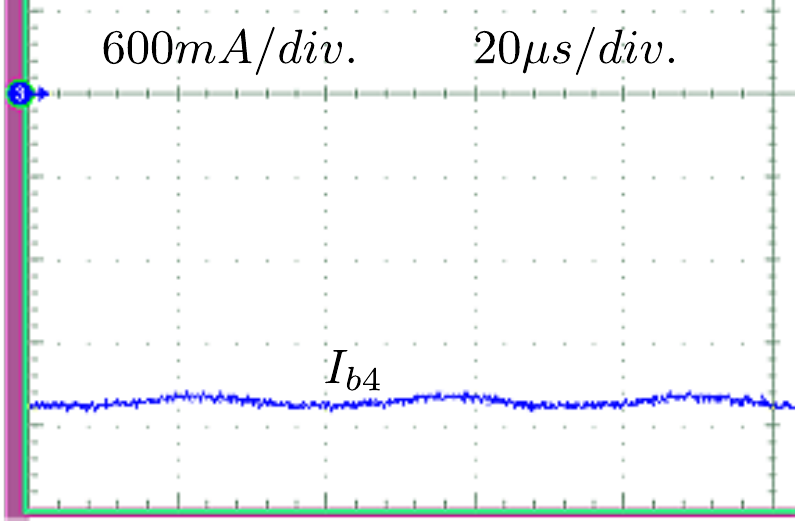}} \end{subfigure}
			
			\caption{Experimental waveforms of (a) battery 1 current, (b) battery 2 current, (c) battery 3 current, (d) battery 4 current when 2 batteries are discharging and two batteries are charging.}
			\label{current_wf1}
		\end{figure}
	\subsection{Testing on a Battery Bank}
	 The developed voltage equalizer circuit is used on a battery bank consisting of four 12 V, 60 Ah lead acid batteries connected in series. In this experiment, gate pulses are generated to discharge batteries 1 and 2 while charging batteries 3 and 4. The Fig.\,\ref{current_wf1} shows the battery currents during this experiment. It can be observed that currents drawn from batteries 1 and 2 are positive, which means that they are discharging. The currents drawn from battery 3 and battery 4 are negative, indicating that they are charging. This verifies that the equalizer is capable of transfer charge from multi-cell to multi-cell.

	\begin{table}[h!]
		
		\centering
		\small
		\caption{Comparison between theoretical and experimental battery currents and powers.}
		\renewcommand{\arraystretch}{1.2}
		\renewcommand{\tabcolsep}{2pt}
		\begin{tabular}{|c|c|c|c|c|c|c|c|c|}
			\hline
			\multirow{3}{*}{} & Battery & \multicolumn{3}{c|}{Battery current} & \multicolumn{3}{c|}{Power}  & G. D.\\ \cline{3-8}
			\multirow{3}{*}{} &  Volt. & Theo. & Exp. & Err. & Theo. & Exp. & Err. & Cur.\\ 
			\multirow{3}{*}{} & ($V$)  &  ($A$) & ($A$) & (\%) & ($W$) & ($W$) & (\%) & ($mA$)\\  \hline
			Bat. 1 & 12.69 &  $2.284$  & $2.427$ & $-5.9$ & $28.98$  & $30.80$ &$-5.9$ & 31 \\ \hline
			Bat. 2 & 12.59 & $2.284$  & $2.356$ & $-3.1$  & $28.76$  & $29.66$ & $-3.1$ & 30\\ \hline
			Bat. 3 & 12.52 & $-2.351$  & $-2.329$ & $0.9$ & $29.43$  & $29.16$ & $0.9$ & 30\\ \hline
			Bat. 4 & 12.04 & $-2.351$  & $-2.305$ & $2.0$ & $28.31$  & $27.75$ & $2.0$ & 29\\ \hline
			\multicolumn{9}{l}{{$^*Note$: Volt: voltage,  Theo: theoretical, Exp: experimental,}} \\ 
				\multicolumn{9}{l}{{Err: error, G.D.: gate driver, Cur: current}}
		\end{tabular}
		\label{results_tab}
		
	\end{table}
	The measured battery currents and powers are given in Table \ref{results_tab} along with their theoretically calculated values using (\ref{ibk}) and (\ref{pk}) which verifies the derived equations. The battery voltage values used in  (\ref{ibk}) and (\ref{pk}) are the measuredd vulues in order to compare with the measured currents and powers.
			
	The inpur power to the equalizer is the sum of powers of discharging batteries (battery 1 and 2) and the output power is the sum of the powers of the charging batteries (battery 3 and 4). The power circuit efficiency of the prototype is calculated to be 94.13\% from the measured battery powers in Table \ref{results_tab}. However, the gate drive loss is not negligible compared to the power level of the circuit of 48 W. Gate drive circuit is powered by the batteries. The total loss in gate drive circuit is calculated to be $1.496W$ from Table \ref{results_tab}. Considering the gate drive loss, the overall efficiency of the voltage equalizer is 91.76\%.

	\begin{figure}[h!]
		\centering
		\begin{subfigure}[]{\includegraphics[width=4cm,height=2.5cm]{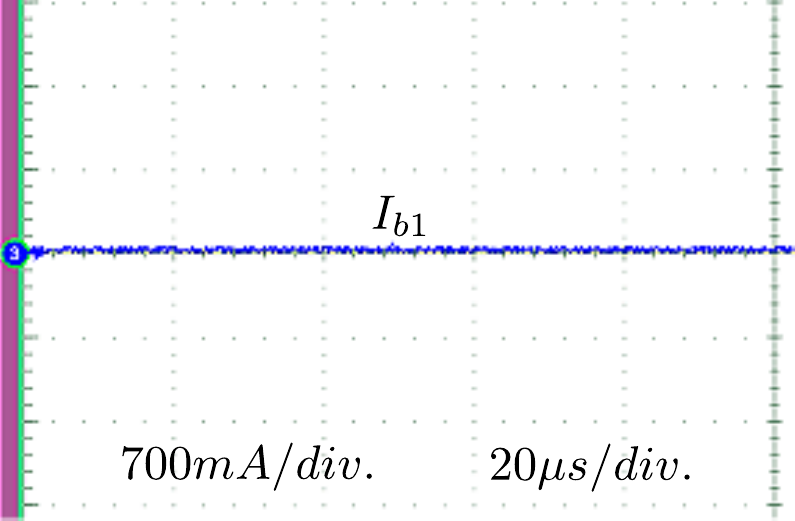}} \end{subfigure}	
		\begin{subfigure}[]{\includegraphics[width=4cm,height=2.5cm]{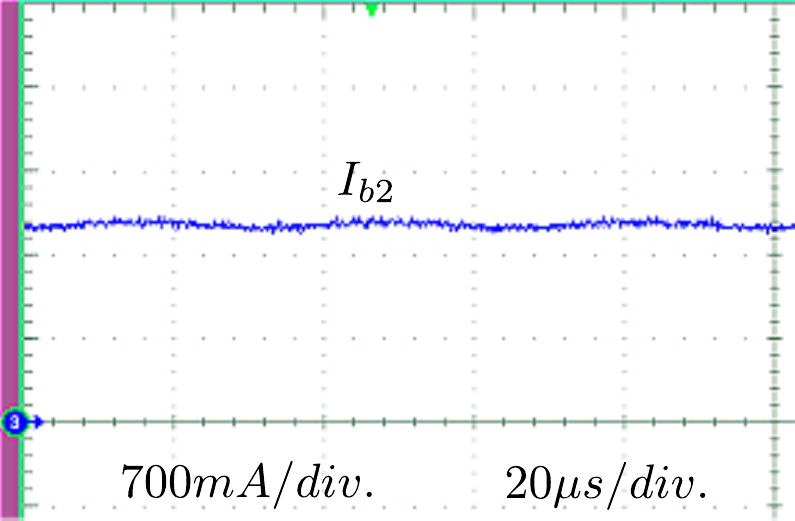}} \end{subfigure}
		\begin{subfigure}[]{\includegraphics[width=4cm,height=2.5cm]{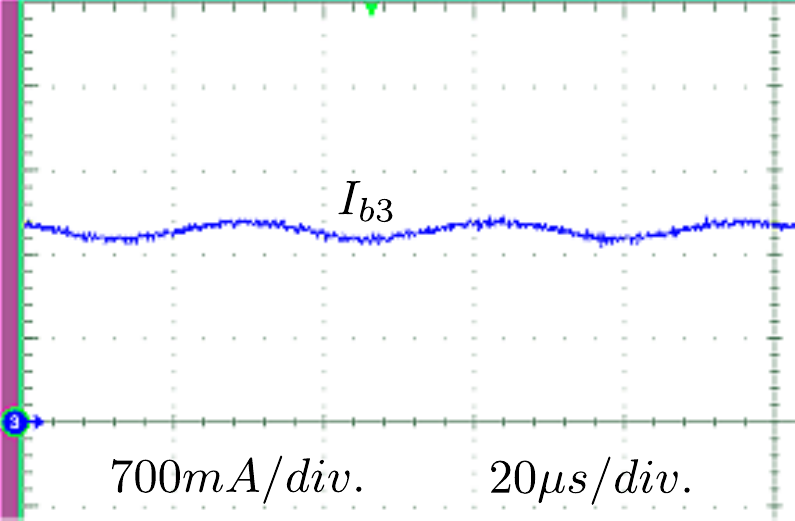}} \end{subfigure}	
		\begin{subfigure}[]{\includegraphics[width=4cm,height=2.5cm]{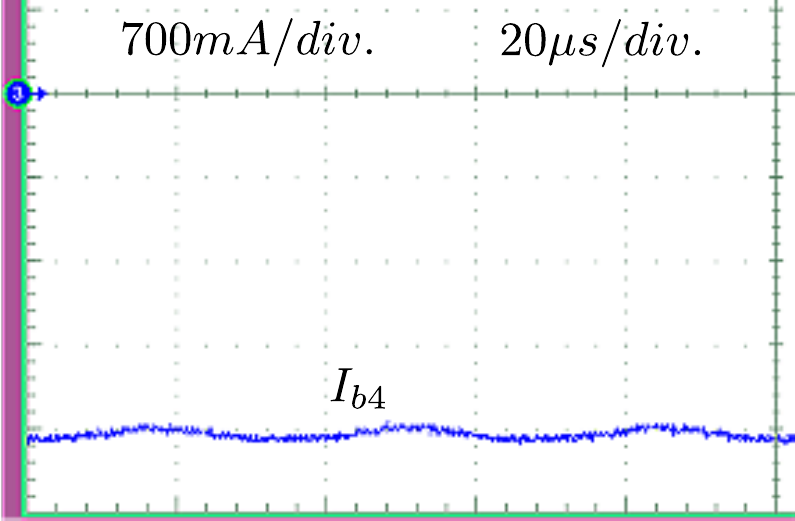}} \end{subfigure}
		
		\caption{Experimental waveforms of (a) battery 1 current, (b) battery 2 current, (c) battery 3 current, (d) battery 4 current when 2 batteries are discharging and one battery is charging.}
		\label{current_wf2}
	\end{figure}

	The battery current waveforms in Fig.\,\ref{current_wf2} are measured when battery 1 voltage is within the acceptable range, battery 2 and 3 need to be discharged and battery 4 needs to be charged. It can be observed from the current waveforms that currents in battery 2 and 3 are positive. The current in battery 4 is negative and is almost twice of the current in battery 2 or 3. Hence, battery 2 and 3 are discharged together to charge battery 4 while current in battery 1 is zero. This observation verifies that the battery whose voltage is within the acceptable range is not charged or discharged while other batteries exchange charge among them and unnecessary charging or discharging is avoided.
	
	\subsection{Voltage Convergence Test}
\begin{figure}[h!]
		\centering
		\includegraphics[width=8.5 cm]{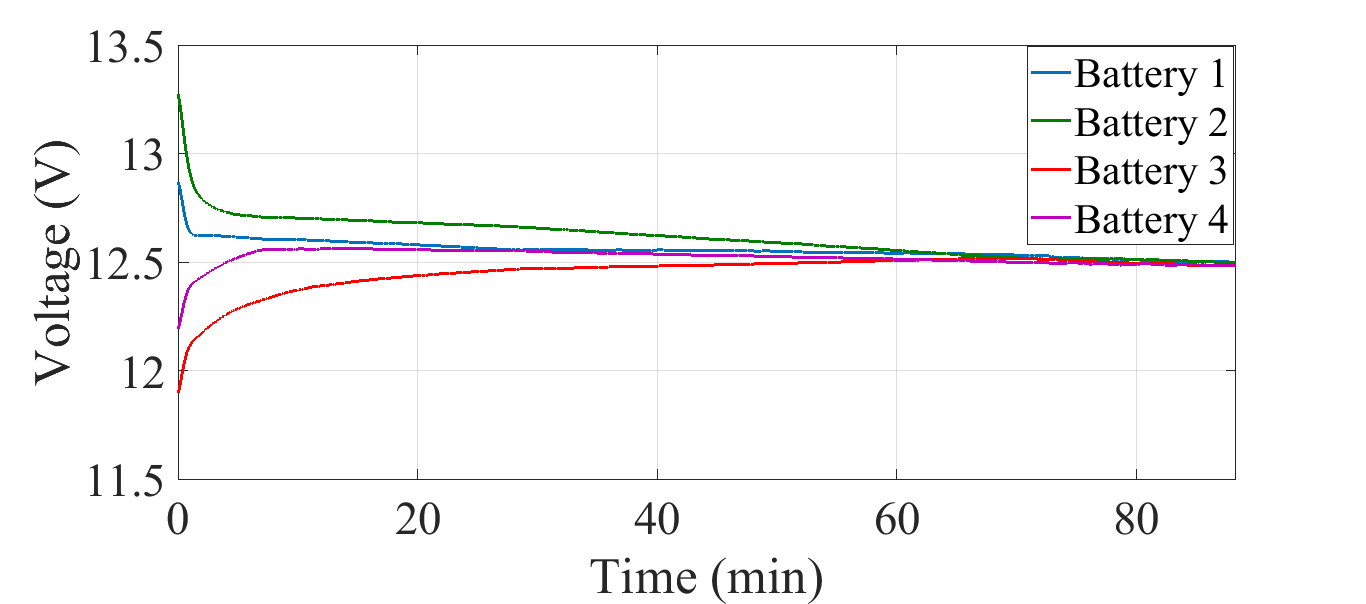}
		\caption{Plot of the battery voltages during voltage convergence test.}
		\label{volt_data}
	\end{figure}

	The equalizer is tested to verify its ability to achieve convergence of the battery voltages. The batteries are unequally charged to create voltage difference among them. Then, the voltage equalizer is used on the battery string and is controlled by the algorithm in Fig.\,\ref{flowchart}. The battery voltages during this test are plotted in Fig.\,\ref{volt_data}. The battery voltages has an initial difference of $1.3V$ among them and converge within a voltage band of $30 mV$ in $90 min$.
	
\subsection{Verification of Soft-switching}
\begin{figure}[h!]
		\centering
		\begin{subfigure}[]{\includegraphics[width=6cm,height=3.0cm]{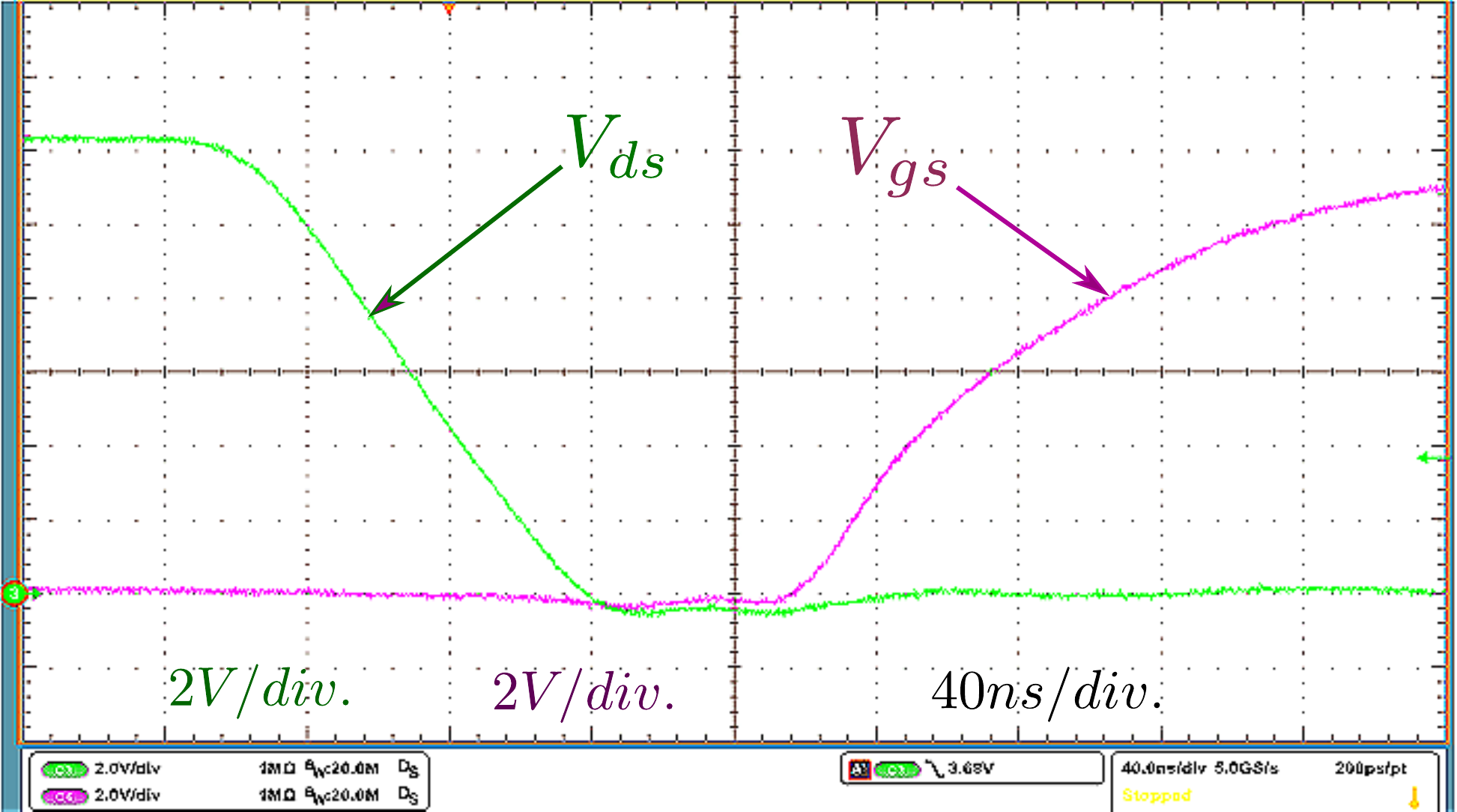}} \end{subfigure}	
		\begin{subfigure}[]{\includegraphics[width=6cm,height=3.0cm]{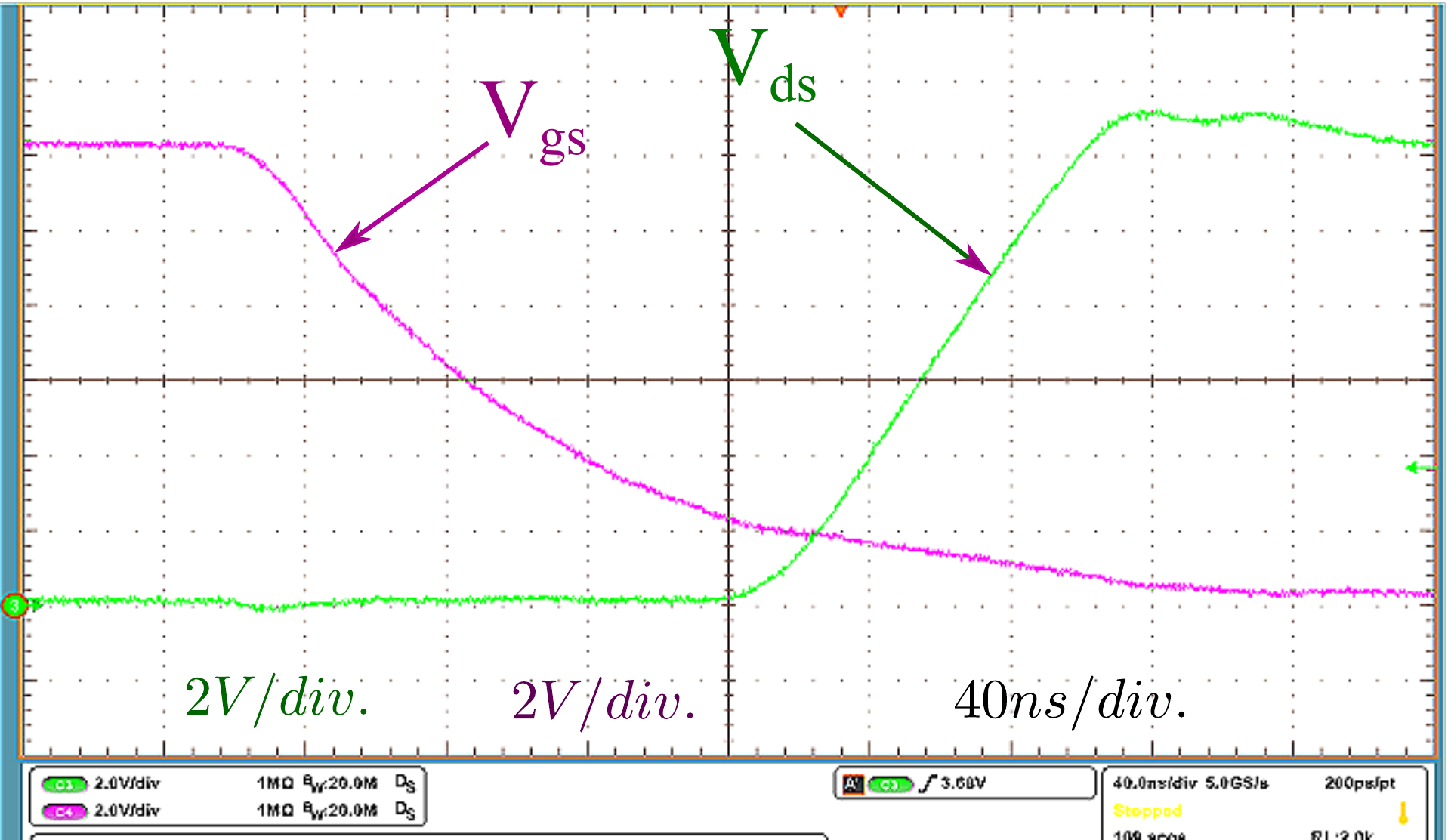}} \end{subfigure}
		
		\caption{Waveforms of drain-source voltage, $V_{ds}$ and gate-source voltage, $V_{gs}$ of a MOSFET during (a) turn-on  and (b) turn-off transitions.}
		\label{exp_sw}
	\end{figure}

	The soft-switching of the proposed topology is also verified with this prototype. The worst condition for achieving soft-switching occurs when the switching current is minimum. As discussed in the appendix, the switching current will be lowest in one of the discharging batteries if only one battery is charging and all other batteries are discharging. In order to verify soft-switching under this condition, switching waveforms are observed when battery 1, battery 2 and battery 3 are discharging while battery 4 is charging. Since the device current cannot be directly measured, the gate-source voltage is measured instead to estimate the current fall and rise in the device. Fig.\,\ref{exp_sw}  shows the drain-source voltage ($V_{ds}$) and gate-source voltage ($V_{gs}$) of the bottom device of the converter connected to battery 1.
	
	Fig.\,\ref{exp_sw}(a) shows the turn-on transition. The voltage across the device, $V_{ds}$ falls to zero before the gate voltage, $V_{gs}$ starts to rise. So, the current rise in the device occurs at zero voltage across the device. Hence, zero voltage switching (ZVS) is achieved in turn-on transition.  Fig.\,\ref{exp_sw}(b)  shows the turn-off transition. The current through the MOSFET reduces to zero when the gate voltage, $V_{gs}$ reduces and crosses the threshold voltage. The threshold voltage of the MOSFET is 2V\cite{mosfet_data}. It can be observed from Fig.\,\ref{exp_sw}(b) that when the gate voltage, $V_{gs}$ falls below 2V, the voltage across the device, $V_{ds}$ is less than 1V. So, the power loss in this transition is very small compared to a hard-switched transition. Hence, ZVS is achieved for turn-off transition as well.
	
%	\begin{figure}[h!]
%		\centering
%		\begin{subfigure}[]{\includegraphics[width=8.5cm]{Results/plot_wo_volt_eq_new.png}} \end{subfigure}
%		
%		\begin{subfigure}[]{\includegraphics[width=8.5cm]{Results/Single_cycle_2A_ckt.png}} \end{subfigure}
%		
%		\caption{Voltages of HUC racks over one charge-discharge cycle (a) without voltage equalizer, (b) with voltage equalizer.}
%		\label{exp_huc}
%	\end{figure}

	\subsection{Testing on a Hybrid Ultra-capacitor Bank}
\begin{figure}[h!]
		\centering
		\includegraphics[width=8.5 cm]{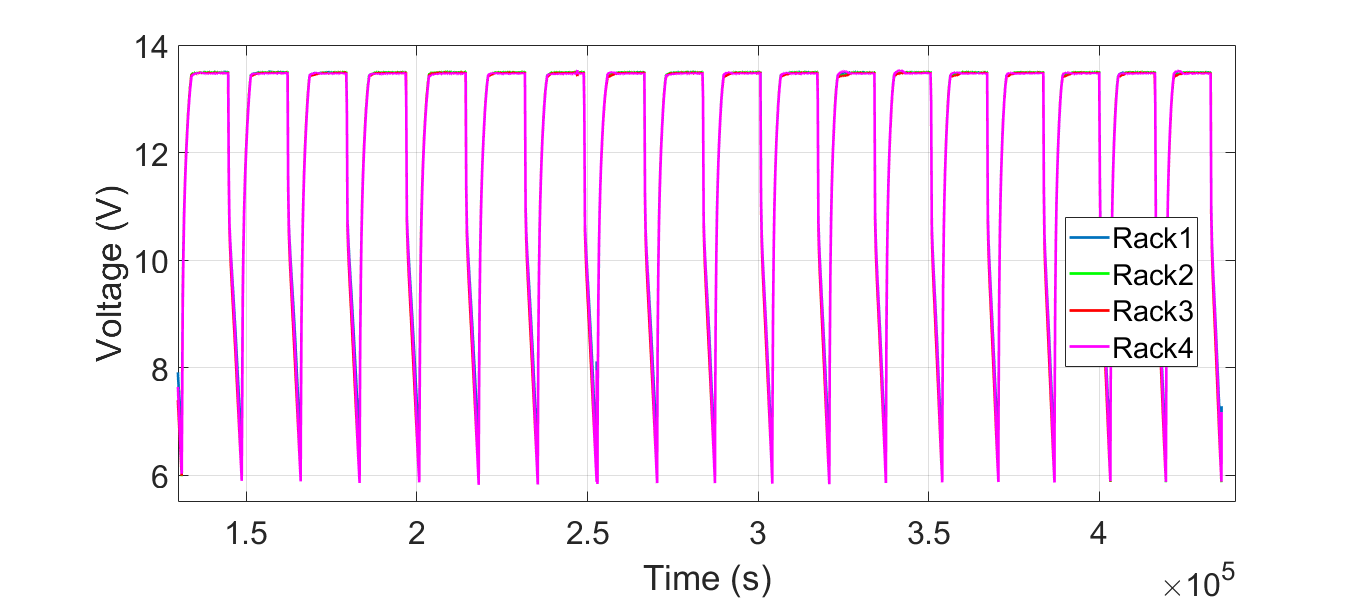}
		\caption{Voltages of HUC racks over 18 charge-discharge cycles with the voltage equalizer.}
		\label{exp_huc_18}
	\end{figure}

	The developed four cell voltage equalizer prototype is used on a hybrid ultra-capacitor (HUC) bank consisting of four HUC racks. Each HUC rack has twenty 12 V, 2500 F HUC\cite{huc} connected in parallel. Four HUC racks are connected in series and the voltage equalizer is used to equalize the voltages of the HUC racks. The series connected HUC racks were charged and discharged by Bitrode Battery Module Test System\cite{bitrode} at 40 A.
%	 Fig.\,\ref{exp_huc}(a) shows the voltages of the four HUC racks over one charge-discharge cycle without any voltage equalizer circuit. Fig.\,\ref{exp_huc}(b) shows the rack voltages with the developed prototype. The effectiveness of the voltage equalizer circuit can be clearly seen by comparing these two plots.
	  Since, HUC takes considerably less time for one charge-discharge cycle, this HUC bank is used to test the equalizer over multiple charge-discharge cycles.
	 The voltage equalizer is tested on the HUC bank continuously over 18 charge-discharge cycles and the recorded voltages are plotted in Fig.\,\ref{exp_huc_18} which shows good performance of the equalizer on the HUC bank with high charge-discharge rate.

	\section{Conclusion}
	A soft switched cell-to-cell voltage equalizer circuit is proposed in this work. This topology avoids unnecessary charging or discharging of any battery to achieve fast voltage equalization. Charge transfer from multiple over-charged batteries to multiple under-charged batteries is achieved simultaneously. A theoretical analysis of the circuit operation is provided which shows that the equalization current does not depend on voltage difference and hence, the equalization speed does not reduce with time. It is shown theoretically that the battery whose terminal voltage is within acceptable range is not unnecessarily charged or discharged. Close loop control of the switches is not required. A simple algorithm for controlling the switching of the converters is provided. The topology operates in ZVS condition. A theoretical analysis of switching current shows that ZVS turn-on is achieved under all battery conditions. ZVS turn-off is achieved with a small capacitor in parallel with each switching device. The turn-off transition is analyzed to determine the value of this capacitor so that ZVS turn-off is achieved under all battery conditions and for which the required dead time can be set for the converter.
	 The circuit operation is verified with a four battery voltage equalizer prototype showing a good match between theoretical estimations and experimentally measured currents and powers. It is also verified experimentally that when switches of a converter is turned off, the corresponding battery is neither charged nor discharged. A voltage convergence test is performed which shows convergence of voltages within $30mV$ tolerance and verifies the effectiveness of the simple equalization algorithm. The prototype is tested on a hybrid ultra-capacitor bank showing good performance over 18 charge-discharge cycles. A power circuit efficiency of 94.13\% and an overall efficiency of 91.76\% is measured.

\appendix

\subsection{Inductor Current}\label{ind_current}
\begin{figure}[h!]
		\centering
		\includegraphics[width=8.5cm]{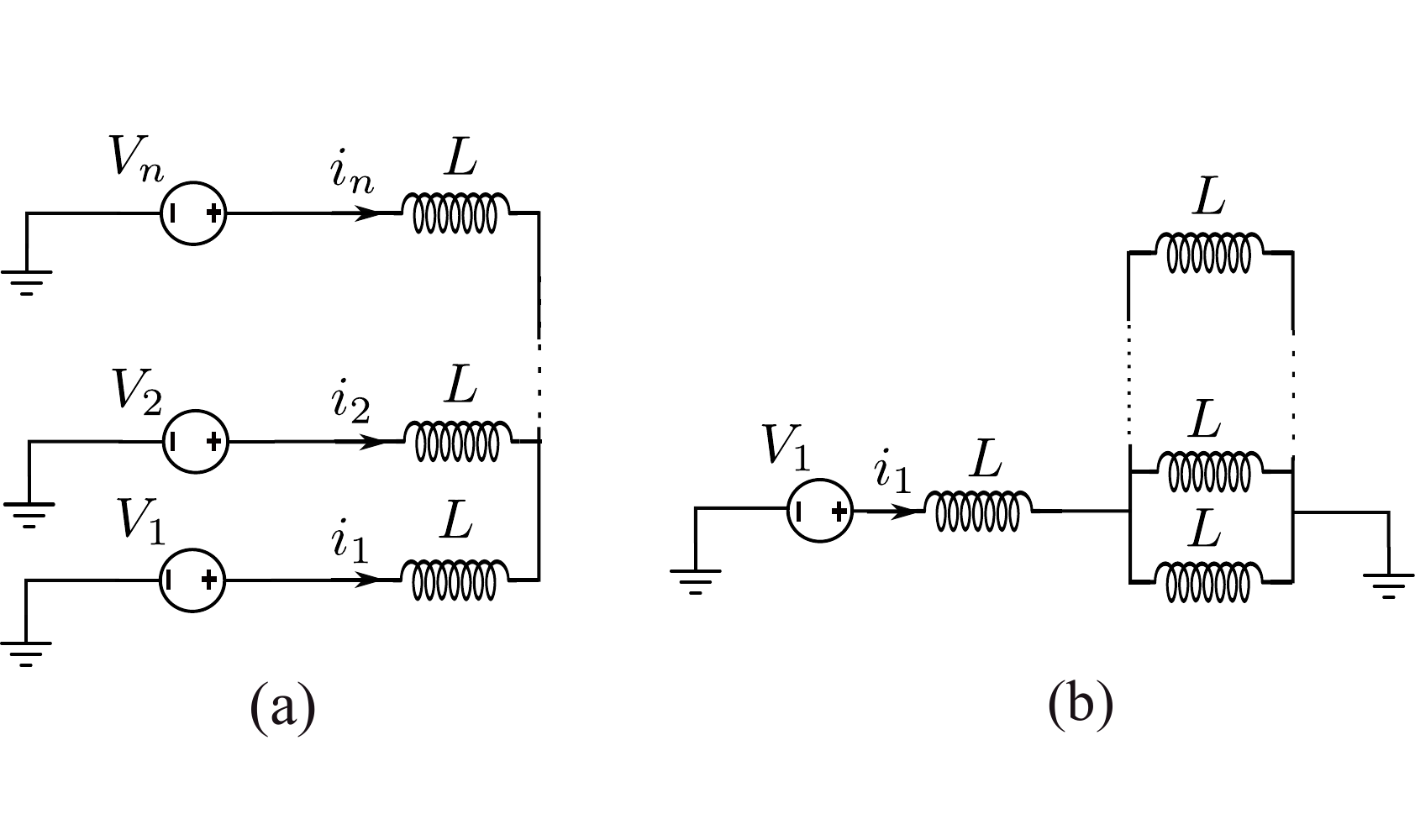}
		\caption{Equivalent circuit for square wave modulation, (a) all the sources are present, (b) all sources except one are replaced by short circuit.}
		\label{eqv_sq}
	\end{figure}

	The pole voltage of the $k^{th}$ converter in Fig.\,\ref{schematic} with respect to the negative terminal of the $k^{th}$ battery is given by (\ref{vpk1}).
	The first term in (\ref{vpk1}) is a dc quantity and the second term is an ac square wave quantity. Since, the capacitor $C$ connected to poles of the converters blocks dc voltage, only the ac component of the pole voltage takes part in power transfer. Hence, the ac equivalent circuit is sufficient for calculation inductor current. 
	The ac equivalent circuit is given in Fig.\,\ref{eqv_sq}(a) where the voltage sources are given by,
	\begin{equation}
	V_k=\frac{V_{bk}}{2}Sq(t+\delta_kT_s) \text{ where }k=1,2,...,n
	\end{equation}

	\begin{figure}[h!]
		\centering
		\includegraphics[width=4.5cm]{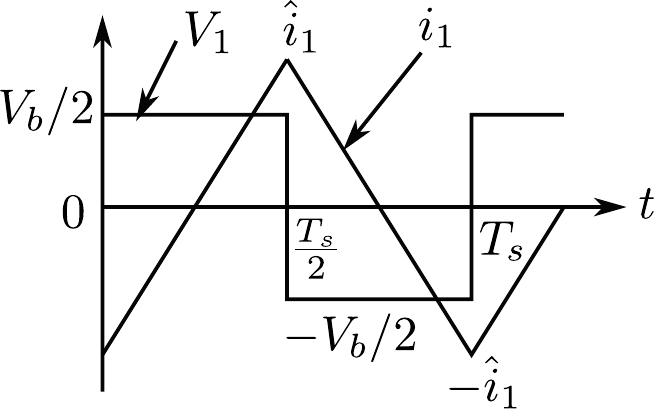}
		\caption{Applied voltage and resulting inductor current when only one source is present.}
		\label{ind_cur}
	\end{figure}

	Principle of superposition is used to obtain the current through each voltage source. The equivalent circuit in Fig.\,\ref{eqv_sq}(b) is obtained by  assuming that only $V_1$ is active and other sources are replaced by short circuit. Equivalent inductance seen by the source $V_1$ is given by,
	\begin{eqnarray}
	L_{eq}=L+\frac{L}{n-1}=\frac{nL}{n-1}
	\end{eqnarray}
	So, the currents drawn from voltage sources in this condition are given by,
	\begin{eqnarray}
	%&=&\frac{V_{b1}T_s}{8L_{eq}}Tr(t+\delta_1T_s)\\
	i_{1}&=& \frac{(n-1)V_{b1}T_s}{8nL}Tr(t+\delta_1T_s)	\\
	i_{i}&=&-\frac{i_1}{n-1}\text{       for i=2,3,...,n}	\\
	&=&	-\frac{V_{b1}T_s}{8nL}Tr(t+\delta_1T_s)
	\end{eqnarray}
	
	Now, considering contributions of all the sources and using superposition,
	\begin{flalign}
	%&=\frac{T_s}{8nL}\left[(n-1)V_{b1}Tr(t+\delta_1T_s)-\sum_{i=2}^{n}V_{bi}Tr(t+\delta_iT_s)\right]\nonumber\\
	i_{1}&= \frac{T_s}{8nL}\left[nV_{b1}Tr(t+\delta_1T_s)-\sum_{i=1}^{n}V_{bi}Tr(t+\delta_iT_s)\right]
	\end{flalign}
	
	In general, the inductor current in the $k_{th}$ converter can be expressed as, 
	\begin{eqnarray}
	\label{ilk1}
	i_{k}=\frac{T_s}{8nL}\left[nV_{bk}Tr(t+\delta_kT_s)-\sum_{i=1}^{n}V_{bi}Tr(t+\delta_iT_s)\right]
	\end{eqnarray}

	 Let's consider that $m$ number of batteries are discharging and $(n-m)$ number of batteries are charging. Then, without any loss of generality, it can be assumed that batteries $1$ to $m$ are discharging while the rest of them are charging. As discussed in Section \ref{modulation}, $\delta_k=0$ for discharging converters and $\delta_k=-\delta$ for all charging converters. Lets consider $k^{th}$ converter, where $k\in (1,m)$. Now, using these gate signal phases in (\ref{ilk1}), the expression for inductor current can be reduced to,
%	\begin{equation}
%	\label{pk_reduced}
%	P_k=\frac{V_{bk}}{4nLf_s}
%	\left[\sum_{i=m+1}^{n}V_{bi}\delta\left(1-2\delta\right)\right]
%	\end{equation}
%	It can be shown that the power expression in (\ref{pk_reduced}) has maximum at $\delta=1/4$ which results in a phase difference of $\pi/2$. Hence, $\delta$ is chosen to be less than $1/4$.  The expression of inductor current $i_k$ in (\ref{ilk}) can be expressed as follows,
%	
	\begin{eqnarray}
	i_{k}&=&\frac{T_s}{8nL}\Bigg[\left((n-1)V_{bk}-\sum_{i=1,i\ne k}^{m}V_{bi}\right)Tr(t)\nonumber\\&&-\left(\sum_{i=m+1}^{n}V_{bi}\right)Tr(t-\delta T_s)\Bigg]\\
	\label{ilk_short}
	&=& c\left[a Tr(t)-b Tr(t-\delta T_s)\right]
	\end{eqnarray}
	Where,
	\begin{eqnarray}
	\label{a}
	a&=&(n-1)V_{bk}-\sum_{i=1,i\ne k}^{m}V_{bi}\\
	\label{b}
	b&=&\sum_{i=m+1}^{n}V_{bi}\\
	\label{c}
	c&=&\frac{T_s}{8nL}
	\end{eqnarray}

	The minimum current in switches during switching transitions determines the minimum required dead-time for ensuring soft-switching. Hence, the inductor current is analyzed in the next subsection to find out its minimum value during switching transitions.
	
	\begin{figure}[h!]
		\centering
		\includegraphics[width=8cm]{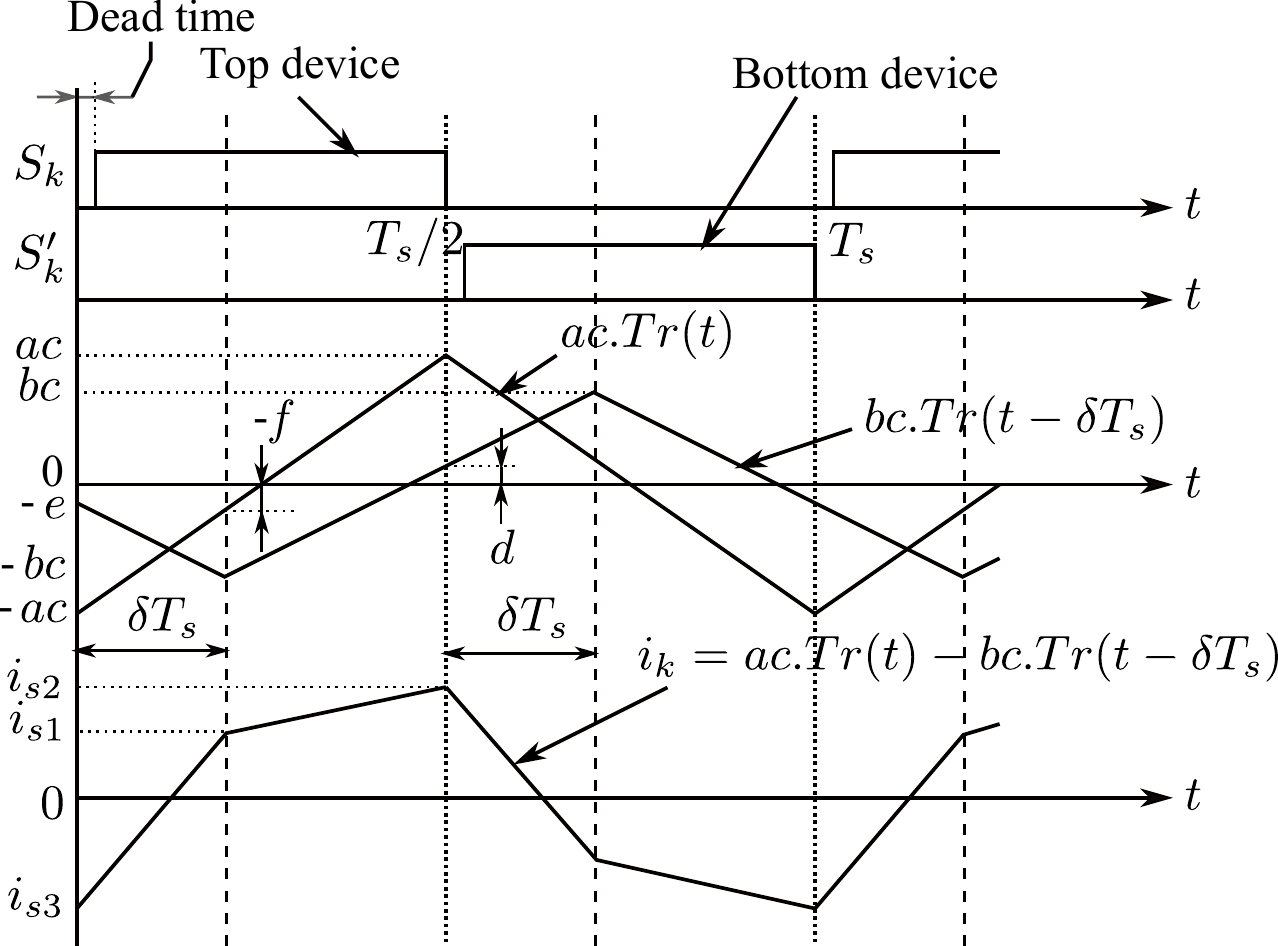}
		\caption{Waveforms of gate pulses and inductor current of a converter leg where the battery is discharging.}
		\label{ind_cur_soft}
	\end{figure}

	\subsection{Upper Limit on $(i_k(t=0))$}\label{min_current}%--------------------------------------------------------------------------------------------------------------
	The inductor current in $k^{th}$ converter is shown in Fig.\,\ref{ind_cur_soft} using (\ref{ilk_short}) along with the gate pulses to top and bottom devices of the converter leg. From the Fig.\,\ref{ind_cur_soft}, the following quantities can be calculated,
	\begin{eqnarray}
	d&=&-bc+\frac{4}{T_s}bc(T_s/2-\delta T_s)=bc(1-4\delta)\\
	e&=&-(-bc+4bc*\delta)=(1-4\delta)bc=d\\
	\label{f}
	f&=&-(-ac+\frac{4}{T_s}ac.\delta T_s)=(1-4\delta)ac
	\end{eqnarray}
The switching instances in a time period shown in Fig.\,\ref{ind_cur_soft} are $t=0$ and $t=T_s/2$ for all converters connected to  discharging batteries including the $k^{th}$ converter.
	%The minimum inductor currents in those instances have to be calculated for finding out the necessary dead-time and value of capacitor $C_s$.
	
	The inductor currents in the $k^{th}$ converter at the time instance $t=0$ is obtained from Fig.\,\ref{ind_cur_soft} as follows,
	\begin{flalign}
	\label{is2}
	&i_k(t=0)=-(ac-d)=-i_{s2}
	\end{flalign} 

	\hspace{0.5cm}

	The expression of $i_k(t=0)$ is derived from (\ref{is2}) using (\ref{a}), (\ref{b}) and (\ref{c}), as follows,
	\begin{eqnarray}
	\label{ib1}
	-i_k(t=0)=c\left[nV_{bk}-\sum_{i=1}^{m}V_{bi}-\sum_{i=m+1}^{n}V_{bi}(1-4\delta)\right]
	\end{eqnarray}
	Here,  $(1-4\delta)>0$ since $\delta<1/4$.  Lets define,
	\begin{eqnarray}
	V_M&=&\frac{\sum_{i=1}^{m}V_{bi}}{m}\\
	V_m&=&\frac{\sum_{i=m+1}^{n}{V_{bi}}}{n-m}\\
	\label{vavg}
	V_{avg}&=&\frac{\sum_{i=1}^{n}V_{bi}}{n}=\frac{mV_M+(n-m)V_m}{n}
	\end{eqnarray}
	Now, from (\ref{ib1}), $V_{bk}$ has to be minimum  to minimize $(-i_k(t=0))$.
	Since, $k^{th}$ battery is discharging, it is overcharged. Hence, the voltage of $k^{th}$ battery must be higher than the average battery voltage of the battery string. So, in worst case,
	\begin{eqnarray}
	V_{bk(min)}=V_{avg}
	\end{eqnarray}
	From (\ref{ib1}),
	\begin{flalign}
	&-i_k(t=0)= c\left[nV_{avg}-\sum_{i=1}^{m}V_{bi}-\sum_{i=m+1}^{n}V_{bi}(1-4\delta)\right]\nonumber\\
	&= c\left[(mV_M+(n-m)V_m)-mV_M-(n-m)V_m(1-4\delta)\right]\nonumber\\
	\label{ib2}
	&= 4\delta c(n-m)V_m
	\end{flalign}
	To minimize the expression of $(-i_k(t=0))$ in (\ref{ib2}), the worst case is considered where, $m=m_{max}=n-1$ and $V_m=V_{bmin}$. The minimum value of $-i_k(t=0)$ is obtained as follows,
	\begin{eqnarray}
	\label{is2min}
	-i_k(t=0)\ge \frac{\delta}{2nLf_s}V_{bmin}\\
	\Rightarrow i_k(t=0)\le -\frac{\delta}{2nLf_s}V_{bmin}
	\end{eqnarray}

%	\subsubsection{Value of $i_{s1}$ when $i_{s2}$ is minimum }
%	Using (\ref{b}), (\ref{c}) and (\ref{f}) in (\ref{is1}), the expression for $i_{s1}$ is obtained,
%	\begin{flalign}
%	\label{is1_new}
%	i_{s1}&=c\left[\sum_{i=m+1}^{n}V_{bi}-(1-4\delta)\left(nV_{bk}-\sum_{i=1}^{m}V_{bi}\right)\right]
%	%&=c\left[\sum_{i=m+1}^{n}V_{bi}+(1-4\delta)\sum_{i=1}^{m}V_{bi}-(1-4\delta)nV_{bk}\right]
%	\end{flalign}
%	
%	
%	The conditions for minimization of $i_{s2}$ are $V_{bk}=V_{avg}=\sum_{i=1}^{n}V_{bi}/n$ and $m=n-1$. Using these conditions in (\ref{is1_new}),
%	\begin{flalign}
%	i_{s1}&=c\left[V_{bn}+(1-4\delta)\sum_{i=1}^{n-1}V_{bi}-(1-4\delta)\sum_{i=1}^{n}V_{bi}\right]\\
%	%&= c\left[\sum_{i=m+1}^{n}V_{bi}-(1-4\delta)\sum_{i=m+1}^{n}V_{bi}\right]\\
%	&=4\delta c V_{bn}
%	\end{flalign}
%	Since, the $n^{th}$ converter is assumed to be charging,  the worst case value of $i_{s1}$ is given by,
%	\begin{eqnarray}
%	i_{s1(min)}&=&4\delta c V_{bn(min)}\\
%	\label{is1min}
%	&=& \frac{\delta}{2nLf_s}V_{bmin}
%	\end{eqnarray} 
%	
%	

	\subsection{Upper Limit on $(-i_k(t=0))$}\label{max_current}
	The expression of $(-i_k(t=0)$ in (\ref{ib1}) is rearranged as follows,
	\begin{eqnarray}
	\label{ib3}
	-i_k(t=0)&=&c\bigg[(n-1)V_{bk}-\sum_{i=1,i\ne k}^{m}V_{bi} \nonumber\\
	&&\qquad-\sum_{i=m+1}^{n}V_{bi}(1-4\delta)\bigg]
	\end{eqnarray}
	The expression in LHS of (\ref{ib3}) is maximized for a given $m$ if $V_{bk}=V_{bmax}$ and $V_{bi}=V_{bmin}$ for  $i\in[1,2,...,m-1,m+1,..n]$, and the following expression is obtained,
	\begin{flalign}
	-i_k(t=0)&=c[(n-1)V_{bmax}-(m-1)V_{bmin}\nonumber\\
	&\qquad \qquad \qquad \qquad  -(n-m)V_{bmin}(1-4\delta)] \nonumber\\
	\label{ib4}
	&=c\bigg[(n-1)V_{bmax}-(n-1)V_{bmin} \nonumber\\
	&\qquad\qquad\qquad+4\delta(n-m)V_{bmin}\bigg]
	\end{flalign}
	The current $-i_k(t=0)$ in (\ref{ib4}) is maximized if $m=m_{min}=1$,
	\begin{eqnarray}
	\label{is2max}
	-i_k(t=0) \le (n-1)c[V_{bmax}-(1-4\delta)V_{bmin} )]
	\end{eqnarray}
	
%	\newpage
	
%	\vspace{10cm}
		\bibliographystyle{IEEEtranTIE}
		\bibliography{IEEEabrv,reference_new} %IEEEabrv instead of IEEEfull

\end{document}